\documentclass[article]{aastex631}
\usepackage[utf8]{inputenc}
\usepackage{natbib, xspace}
\usepackage[caption=false]{subfig}

\usepackage{amsmath}
\date{\today}

\shorttitle{Heterogeneous Enrichment from Primordial Stars}
\shortauthors{Wells and Norman}

\newcommand{\msun}{${\rm M}_\odot$\xspace}
\newcommand{\enzo}{{\tt Enzo}\xspace}
\newcommand{\enzoe}{{\tt Enzo-E}\xspace}
\newcommand{\z}{\text{[Z/H]}}

\newcommand{\R}{M_{Z*}/M_{ZH}}

\newcommand{\pa}{PIII association\xspace}
\newcommand{\pas}{PIII associations\xspace}

\newcommand{\age}{a_{\rm Myr}\xspace}

\newcommand{\starss}{{\tt STARSS}\xspace}
\newcommand{\starnet}{\texttt{StarNet}\xspace}
\newcommand{\starfind}{\texttt{StarFind}\xspace}
\newcommand{\piii}{Population III\xspace}
\newcommand{\pii}{Population II\xspace}
\begin{document}

\title{The First Galaxies and the Effect of Heterogeneous Enrichment from Primordial Stars}

\author{Azton Wells}
\email{aiwells@ucsd.edu}
\affil{Center for Astrophysics and Space Sciences\\
University of California, San Diego, La Jolla, CA, 92093}
\author{Michael L. Norman}
\affil{Center for Astrophysics and Space Sciences\\
University of California, San Diego, La Jolla, CA, 92093}
\affil{San Diego Supercomputer Center\\
University of California, San Diego, La Jolla, CA, 92093}
\begin{abstract}
We incorporate new scale-intelligent models of metal-enriched star formation (\starss) with surrogate models of primordial stellar feedback (\starnet) into the astrophysics simulation code \enzo to analyze the impact of heterogeneous metal enrichment on the first galaxies.  Our study includes the earliest generations of stars and the protogalaxies ($10^6 \lesssim M_v/M_\odot \lesssim 10^8$) containing them. We compare results obtained with the new methods to two common paradigms of metallicity initial conditions in simulations: ignoring the metallicity initial condition and assuming a uniform metallicity floor.  We find that ignoring metallicity requirements for enriched star formation results in a redshift-dependent excess in stellar mass created and compounding errors consisting of stars forming in pristine gas. We find that using a metallicity floor causes an early underproduction of stars before $z=21$ that reverses to overproduction by $z=18$.  At the final redshift, $z=14.95$, there is $\sim 20\%$ excess stellar mass with 8.6\% increased protogalaxy count. Heterogeneous metallicity initial conditions greatly increase the range of halo observables, e.g., stellar metallicity, stellar mass, and luminosity.  The increased range leads to better agreement with observations of ultra-faint dwarf galaxies when compared to metallicity-floor simulations.   \starnet generates protogalaxies with low stellar mass, $M_* \lesssim 10^3 M_\odot$, so may also serve to model low-luminosity protogalaxies more effectively than a metallicity floor criterion at similar spatial and mass resolution.

\end{abstract}
\section{Introduction}
Historically, astrophysical simulations have been restricted to single, purpose-motivated scales, e.g, using gravity only or gravity+hydrodynamics simulations to study cosmology \citep{vogels2013, vogels2014, emberson2019}, increasing resolution and adding sub-grid star formation and feedback recipes to study single galaxies \citep{hopkins2018, wheeler2019, emerick2020}, or monumentally increasing the resolution and including recipes for the formation and feedback of the first stars to study individual star forming mini-halos \citep{wise2011, wise2012, smith2015}.  

Because of the extreme difference in scale between star formation and large-scale structure formation, even next generation simulation codes running on exascale systems will struggle to comprehensively model star formation and feedback in a large-scale cosmological volume.    For example, in the \textit{Phoenix Simulations}(PHX) \citep{wells2022}(W22), $>50\%$ of simulation time is spent evolving adaptively refined high-resolution regions required for \piii star formation and feedback. This problem is further exacerbated by evolving their associated supernova remnants (SNR) at an early and hot stage at high resolution.  In addition, both the \piii and \pii stellar feedback routines used in the PHX are \textit{not} scale insensitive--they deposit supernova (SN) energy in thermal form, which places severe limits on spatial resolution \citep{martizzi2015, rosdahl2016}. Taking full advantage of next-generation exascale systems will require not only massively parallel simulation codes, but novel accelerations within those codes.  In this work, we attempt to lower the extreme spatial and temporal resolution requirements of prior works as a method of acceleration. 

The problems outlined above suggest two major avenues of progress for next-generation simulations in \enzo\footnote{https://enzo.readthedocs.io} and its massively parallel, CHARM++-based successor, \enzoe\footnote{https://enzo-e.readthedocs.io/en/main/}\citep{bordner2018}. First, since the current methods of metal-enriched star formation in \enzo are very sensitive to spatial resolution, we require a scale-intelligent star formation and feedback algorithm.  Motivated by other recent work using algorithms that adapt to resolution in some sense \citep{kimm2014,rosdahl2017,hopkins2018} , we have developed the \textbf{S}cale-intelligent \textbf{T}erminal-momentum \textbf{A}lgorithm for \textbf{R}ealistic \textbf{S}tellar \textbf{S}ources (\starss).  Using physically motivated parameterizations and prior simulation results, \starss uses a minimum of user-defined parameters to implement star formation and feedback that does not require tuning by the user prior to their next simulation.  In Section \ref{sec:starformation}, we describe the \starss algorithm, while in Section \ref{sec:testing} we describe how the model performs in testing at various resolutions.  

Second, we address the issue of \piii star formation and feedback.  Despite the fact that \piii star formation is not uniform or isotropic, many current works assume a uniform prior distribution of metals \citep[e.g.,][]{hopkins2018}, thereby negating the need for any model to address the \piii era.  Other works simply ignore the first generation completely, allowing enriched star formation to occur in pristine gas \citep[e.g.,][]{vogels2014}.  Both methods would negate the effect of \piii stars, e.g., prompt star formation from powerful SN \citep{machida2005, ritter2012, chiaki2013, wells2022}, destruction of early halos from \piii SN events \citep{whalen2008}, and the possibility of halos that go unenriched by \piii star formation \citep{regan2017}.  This work combines prior efforts in \cite{wells2021}(W21) and W22 to create a deep learning accelerated surrogate model that addresses the non-uniformity of halo enrichment from \piii SNe. The framework developed for this purpose is referred to as the { StarNetRuntime} or simply {\tt StarNet}.  Although we developed it for the purpose of modelling \pas\footnote{As noted in W22, this term refers to a group of coeval \piii stars, too small to form a canonical cluster, but too large to be a simple binary/trinary system.}, the paradigms of \starnet would be readily adaptable to modelling single metal-enriched stars, metal-enriched star clusters, or even galaxy level star formation and feedback.  \starnet is described in Section \ref{sec:starnet}, while Section \ref{sec:starnet_comps} describes the results of \starnet simulations and compares them to simulations with a metallicity floor using \starss 
in order to quantify the impact of intelligently heterogeneous enrichment on the first protogalaxies and their minihalos. 

\section{\textbf{S}cale-intelligent \textbf{T}erminal-momentum \textbf{A}lgorithm for \textbf{R}ealistic \textbf{S}tellar \textbf{S}ources (\starss)}
\label{sec:starformation}
    The primary motivation of this work is to somewhat relax the severe resolution requirements of simulations of the first galaxies, such as the \textit{Birth of a Galaxy} \citep{wise2012}, while maintaining as much of the physical effects of the star formation and feedback models as is possible.  Currently, there is no resolution-insensitive method in \enzo; both the Population II star cluster and Population III single-star models have resolution sensitive criteria for star formation requiring specific overdensities for star formation, however the stellar feedback is also extremely resolution sensitive.  All SN or wind feedback from either stellar source is placed into a sphere of $\sim 10$ pc as an approximation of the Sedov-Taylor phase of supernova remnant (SNR) \citep{sedov1946propagation, taylor1950formation}.  This particular paradigm of stellar feedback is known to require strict resolution requirements, such as resolving the feedback radius by $>4$ cells \citep{kim2015, simpson2015, rosdahl2016, hopkins2018}.  With motivation from these prior works, we develop a stellar feedback algorithm in this section that is less sensitive to changing resolution (resolution invariance is sadly still out of reach).
\subsection{Star Particle Formation}
Star particle formation follows a very similar prescription as described in \cite{hopkins2018}(FIRE-2), and can optionally follow the updated methods of \cite{hopkins2022}(FIRE-3).  Those criteria that can be ignored or modified to follow FIRE-3 are denoted with a ``*" below.  Several criteria are checked at each grid cell to see if that cell qualifies for star formation:
\begin{enumerate}
    \item *$\rho / \bar\rho > \rho_{c}$, where $\rho_{c}$ is the minimum overdensity relative to the simulation volume that is allowed star formation and $\bar\rho$ is the simulation volume average density. Alternatively, the density parameter can consider number density: $n_b > n_{b,c}$.
    \item Converging gas flow: $\nabla\cdot \boldsymbol{v}_{cell} < 0$ for $\boldsymbol{v}_{cell}$, the cell-centered gas velocity. 
    \item The virial parameter ($\alpha$), or ratio of kinetic + internal energy to gravitational potential energy ($E_{g}$) is checked using Enzo's total energy field (available since Enzo uses a dual-energy formalism to track the energy of the gas).  We require $\alpha < 1$ with 
        $$ \alpha = \frac{E_{\rm total} }{E_{\rm g} }, $$
    where all quantities refer to the cell-centered value.  Canonically, $\alpha = 2 E_k / E_g$ for the kinetic energy, $E_k$ \citep{mckee1992}.  Here, we use $E_{\rm total}$ to explicitly include contributions from all energy sources.
    \item The cooling time must be less than the freefall time: $t_c < t_{\rm ff}$ with $t_{\rm ff} =  \sqrt{(3\pi)/(32 G \rho)}$, or temperature $T< 10^4$ K.
    \item The gas mass in the cell must be greater than the critical Jeans mass: $m_b > \max(m_{j}, 10^3$M$_\odot$).
    \item* The gas must have self-shielded hydrogen fraction, $f_s > 0$.  This is checked via the analytic approximation of \cite{krumholz2011}.  If using 9-species chemistry, this requirement can also be explicitly checked by comparing neutral to ionized molecular hydrogen as evolved in the \enzo chemistry routines.
    \item The metallicity must be above some user-defined critical value, $Z_{c}$ \footnote{$Z$ is defined as the log of metal abundance relative to solar metallicity; for metal mass $M_z$ and baryon mass $M_b$, $Z = \log(M_z/M_b) - \log(M_{z,\odot}/M_\odot)$}. Alternatively, the metallicity threshold can be ignored, although it is still included in the analytic approximation of $f_s$. 
\end{enumerate}
If the above relevant criteria are satisfied, the creation routine assigns an integrated star formation rate 
\begin{equation}
    \dot M_* = \frac{f_s \eta_{\rm sf} M_b }{ t_{\rm ff}},
\end{equation} 
with maximum allowed star formation efficiency\footnote{This efficiency is simply the maximum fraction of gas within a cell that would be allowed to convert to stars within a timestep; this avoids converting all gas to stars and generating simulation failures from the resulting discontinuity in the density field. It does not imply a global efficiency, e.g., of gas to star conversion within a galaxy.} $\eta_{\rm sf}$.  The probability of star formation is then given for the grid timestep $\delta t$ as
\begin{equation}
    p_{\rm sf} = 1-\exp\bigg[-\frac{\dot M_*\delta t}{\eta_{\rm sf}M_b}\bigg].
\end{equation} A star particle is formed only if sampling a binomial distribution with $p_{\rm sf}$ yields success.  With probability  satisfied, a new star is formed with $M_* = \min(f_s \eta_{\rm sf} M_b, M_{\rm *,max})$, where  $M_{\rm *,max}$ is an optional user-defined maximum star mass.  The user may specify no maximum mass, in which case $M_* = f_s \eta_{\rm sf} M_b$.  However, if $M_*$ is large, there may be times where one would expect $>1$ SNe per particle in a timestep.  Such a situation is undesirable, since \starss feedback is designed to couple single SN events to the grid and its analytic models do not in general hold true for combined events that would result in sub-grid ``super-bubble" remnants (SBRs).  To avoid many SNe per timestep, we split large (parent) particles into several smaller sub-cluster (child) particles.  Each child maintains the metallicity of the parent with velocity randomly assigned so that the child velocity is $0.95<|{\mathbf v}_p|<1.05$, with the parent velocity $|\mathbf{v}_p|$.  Likewise, we assign the position of the child particle randomly within the same grid-cell as the parent.  Finally, the creation time of the child ($t_c$) is offset from the parent by factors of the dynamical time: for the $i^{th}$ of $n$ sub-clusters flagged for creation at time $t_{c,0}$, the modified creation time is set as $t_{c,i} = t_{c,0} + 3i \times t_{\rm ff}/n~{\rm Myr}$ so that the creation of all sub-clusters is distributed across three $t_{\rm ff}$ \citep{murray2011}.


\subsection{Star Particle Feedback}
\label{sec:feedback}
\subsubsection{Determination of Feedback Quantities}
    This section details the feedback of \starss particles including rates of SN, winds, luminosity and how those rates are used within our implementation.  Except where otherwise noted, the rates in this section are adopted from FIRE-2 which were generated using Starburst 99~\citep{leitherer1999} simulations. For each formed particle, at each timestep, we calculate the age-based rates for supernovae (type II ($R_{\rm ii}$ and type Ia ($R_{\rm ia}$)) and winds.  The probability ($P_x$) of a SN of type $x$ is then $P_x = M_* R_x d\delta t,$ for $\delta t$ being the timestep measured in Myr.  If sampling a binomial distribution with $P_x$ returns success, then a SN event will be modeled for this timestep.
    
    There are three modes of feedback coupled from star particles to the computational grid: Supernovae, winds, and radiation.  Supernovae rates are approximated by piecewise functions that depends on the age of the particle measured in Myr ($\age$):
    
    \begin{equation} R_{\rm ii} = \begin{cases}
        0 & \age < 3.401 \\
        5.408\times10^{-4} & 3.401 \leq \age < 10.37 \\
        2.516\times10^{-4} & 10.37 \leq \age < 37.53 \\
        0 & \age \geq 37.53
        \end{cases}
    \end{equation}
    \begin{equation}
        R_{\rm ia} = \begin{cases}
        0 & \age < 37.53 \\ 
        5.2\times 10^{-8} + 1.6\times10^{-5} & \\ 
            \times\text{exp}\bigg\{-\bigg(\frac{[(\age - 50 )/10]^2}{2}\bigg)^2\bigg\} & \age \geq 37.53\\
        \end{cases}
    \end{equation}
    
    If either $P_{\rm ii}$ or $P_{\rm ia}$ result in a SNe event, we assign an ejecta mass of 10.5 or 1.5 M$_\odot$ to Type-II and Type-Ia SN respectively.  The metal ejecta for Type-II is $M_{z,{\rm ej}}  = 1.91 + 0.0479 \times \max(Z_*, 1.65)$ or 1.4 M$_\odot$ for Type-Ia.  
    
    At each timestep for the particle we additionally derive mass, energy and metal from stellar winds that must be coupled to the grid.  The wind mass is given by $M_{\rm w} = M_* f_{\rm w} \delta t$ Gyr$^{-1}$, with the wind loading factor
    \begin{equation}
        f_{\rm w} = \begin{cases}
            4.763\times\min(0.01+Z, 1)                                   & \age < 1\\
            4.763\age^\kappa\times\min(0.01+Z, 1.0)    & 1 \leq \age < 3.5\\
            29.4\bigg(\frac{\age}{3.5}\bigg)^{-13/4}+0.0042      & 3.5 \leq \age <100\\
            0.43\bigg(\frac{(\age/100)^{-1.1}}{19.81/\log(\age)}\bigg) & 100 \leq \age\\
        \end{cases}
    \end{equation}
    \begin{equation}
        \kappa = {1.45+0.08\times\min(Z, 1)}. 
    \end{equation}
    The mass of metals in the wind, $M_{z,{\rm w}}$ is then given by 
    \begin{equation}
        M_{z,{\rm w}} = \max(0.02, 0.016+ f_z)\times M_{\rm w},
    \end{equation} 
    with
    \begin{equation}
        f_z = 0.0041 \times (\max(Z, 1.65)+0.0118).
    \end{equation}
    Finally, the energy is determined as $E_{\rm w} =10^{12} M_{\rm w} \epsilon$, with $\epsilon$ is defined as:
    $$
        \epsilon = 4.83+
            \frac{5.94\times10^4}{1+(\age/2.5)^{1.4}}+\bigg(\frac{\age}{10}\bigg)^5 
    $$
    for $a_{\rm Myr} < 100$, or $\epsilon = 4.83$ for all other times.
    
    The last form of feedback associated with these star particles is via radiation.  Each particle produces ionizing radiation according to 
    $$\Psi_{\rm ion} = \begin{cases}
            500 & \age < 3.5\\
            60 \big(\frac{\age}{3.5}\big)^{-3.6}+460\big(\frac{\age}{3.5}\big)^\gamma) & 3.5 \leq \age < 25\\
        \end{cases}
    $$
    $$\gamma = 0.045-1.82\log(\age)$$
    with $\Psi$ in units L$_\odot$/M$_\odot$.  With this parameterization, each particle will emit $\sim 8.94 \times 10^{60}$ ionizing photons/M$_\odot$ throughout its 25 Myr radiative lifetime. While SN and winds are coupled to the grid as prescribed in as follows in Section~\ref{sec:coupling}, radiation is coupled directly to the Moray ray-tracing radiation solver \citep{wise2011} in \enzo.

\subsubsection{Coupling Feedback}
\label{sec:coupling}
Given an event from supernovae or stellar winds with ejecta energy $E_{\rm ej}$, mass $M_{\rm ej}$ and metal mass $M_{z,{\rm ej}}$, we couple the feedback to the computational domain as described in this section.  We wish to separate $E_{\rm ej}$ into thermal and kinetic components based on physically motivated analytic expressions: further, we wish to distribute the kinetic energy via explicitly coupling momentum to the gas neighboring the star particle in an isotropic manner. Coupling the momenta in physically meaningful ways implies altering the amount of momenta coupled given the stage of the SNR, which is determined by the grid resolution and locally averaged grid field quantities.  To determine the phase of the SNR, we compare the grid resolution, $dx$, to various quantities: the free expansion radius ($R_{\rm free}$), the cooling radius ($R_{\rm c}$), and the fading radius, ($R_{\rm fade}$).  $R_{\rm free}$ is determined as in \cite{kim2015} as
    \begin{equation}
        R_{\rm free} = 2.75 \bigg(\frac{M_{\rm ej}}{3 M_\odot}\bigg)^{1/3} n_b^{-1/3}\textrm{ pc}.
    \end{equation}
$R_{\rm c}$ is given in FIRE-2, 
    \begin{equation}
        R_{\rm c} = 28.4 n_b^{-3/7}E_{51}^{2/7}f(Z) \textrm{ pc}
    \end{equation}
with $E_{51} = E_{\rm ej} / (10^{51}$ erg) and $f(Z) = Z^{-0.14}$ if $Z\geq 0.01$; else $f(Z) = 2$, because there is little to no metallicity dependence below $Z\sim 0.01$ \citep{thornton1998}.
Finally, $R_{\rm fade}$ represents the radius at which we would expect the SNR to have merged with the ISM, i.e., the expected velocity of the shell is comparable to the local speed of sound.  The expression is taken from \cite{draine2011} as
\begin{equation}
    R_{\rm fade} = 66.0 ~E_{51}^{0.32}n_b^{-0.37}\bigg(\frac{c_s}{10 \textrm{ km s}^{-1}}\bigg)^{-0.4} \textrm{ pc}.
\end{equation}
Since this simplified expression has no consideration for local metallicity, we finally take $R_{\rm fade} =  \min(R_{\rm fade}, 1.5\times R_{\rm c}$).  In all the prior expressions, for SNRs, the energy is taken as $10^{51}$ erg, the mass is 10.5 and 1.5 $M_\odot$ for Type-II and Type-Ia SNe respectively.  

After determining the phase of SNR, we calculate the expected momentum to couple ($p_{cpl}$) using the following expressions:
\begin{equation}
    p_{\rm cpl} = \begin{cases}
            \sqrt{2 M_{ej} E_{ej}}\times (dx / R_{\rm free})^3 & dx < R_{\rm free}\\
            \min(p_{ST}, p_T \times dx/R_{\rm c}) & R_{\rm free} < dx < R_{\rm c}\\
            p_T & R_{\rm c} < dx < R_{\rm fade} \\
            p_T \times \eta& dx > R_{\rm fade}
    \end{cases}
\end{equation}
where $\eta = \big(1.0 - \tanh{[(1.25\times dx/R_{\rm fade})^{2}]} \big)$ to smoothly connect the terminal phase to the fading phase where coupled momentum is zero. The expected momentum of the Sedov-Taylor(ST) phase ($p_{\rm ST}$) at the radius given by $dx$ is
\begin{equation}
    p_{\rm ST} = 2.21\times10^4~E_{51}^{0.8}n_b^{0.2}t_3^{0.6} \textrm{M$_\odot$ km s}^{-1}
\end{equation}
where we solve for $t_3$ taking the radius as $dx$:
\begin{equation}
    t_3 = \bigg(\frac{dx}{5\textrm{ pc} (E_{51} / n_b) ^{1/5}}\bigg)^{5/2}
\end{equation}
as derived from \cite{kim2015}.  The momentum in the momentum-driven snowplough phase (terminal phase, $p_T$) is taken from \cite{thornton1998} as
\begin{equation}
    p_{\rm T} = \begin{cases} 
        1.67\times10^5 E_{51}^{13/14} n_b^{-1/4} Z^{-0.36} & Z > 0.01\\
        8.36\times10^5E_{51}^{13/14} n_b^{-1/4} & \textrm{else}
    \end{cases}
\end{equation}
Finally, \cite{cioffi1988} notes that there is a low-density regime where there is no expected shell formation and the remnant merges with the ISM before the shell mass approaches the ejecta mass.  This case is extremely important in our model, as the SNe originate from a single point in space; after the first few events, the interior region is very hot ($T \geq 10^7$ K) and low-density ($n_b \lesssim 10^{-2}$ cm$^{-3}$).  To smoothly connect the phase of the SNR to this regime where we expect no momentum coupling, in cases where $n_b < 10 n_{\rm C}$ we modify the coupled momentum as
\begin{equation}
    p_{\rm cpl} = p_{\rm cpl} \times \bigg(1 - \tanh\bigg[\big(1.45\frac{n_{\rm C}}{n_b}\big)^{6.5}\bigg]\bigg)
\end{equation}
with critical density 
\begin{equation}
    n_{\rm C}= 0.0038 \frac{(P_4)^{7/9} (v_s/c_s)^{14/9}} {E_{51}^{1/9}} Z^{1/3}
\end{equation}
for $P_4 = n_b T / 10^4$ K and the expected shell velocity $v_s$. In order to calculate $v_s$, we utilize the following expressions for the mass of the SNR shell ($M_s$):
\begin{equation}
    \frac{M_{s}}{\textrm{ M$_\odot$ }} = \begin{cases}   
        M_{\rm ej} & dx < R_{\rm free}\\
        M_{\rm ej}+\frac{4}{3}\pi dx^3 \bar\rho & R_{\rm free} < dx < R_{\rm c}\\
        M_{\rm ej}+1.41\times 10^4  \frac{E_{51}^{6/7}Z^{0.27}}{n_b^{0.24}}  & R_{\rm c} < dx, Z > 0.01 \\
        M_{\rm ej}+4.89\times 10^4 \frac{E_{51}^{6/7}}{n_b^{0.24}}  & R_{\rm c} < dx, Z < 0.01 \\
        
    \end{cases}
\end{equation}
The first case assumes no shell mass in the free expansion stage.  The only mass coupled in this stage is the ejecta mass.  The second assumes that the shell mass behaves as if sweeping up mass in a sphere with averaged density $\bar\rho$\footnote{with $\bar\rho$ as the average density in a $3^3$ patch centered on the feedback source.}.  The final two expressions describe the shell mass in the terminal phase,  at different metallicity regimes~\citep{thornton1998}.  Of course, these analytic expressions have to consider the mass that exists on the grid; we determine the mass that exists within the central cells ($M_{\rm central}$) where the shell mass would be removed and the final shell mass is limited to $M_s = \min (0.75 M_{\rm central}, M_s)$. With $M_s$ in hand, the velocity of the shell is given by $v_s = p_{\rm cpl}/M_s$ and the metal contribution from the shell is given as the mass averaged metallicity of the central cells from which the shell is being evacuated.

In principle the analytic forms of this section are also applied to stellar winds.  In practice however, the wind energy is so low that the momentum is negligible.  Despite the lack of momentum coupling at lower resolutions, the mass, gas energy and metal from winds are still deposited.

\begin{figure}
    \centering
    \includegraphics[width=0.48\textwidth]{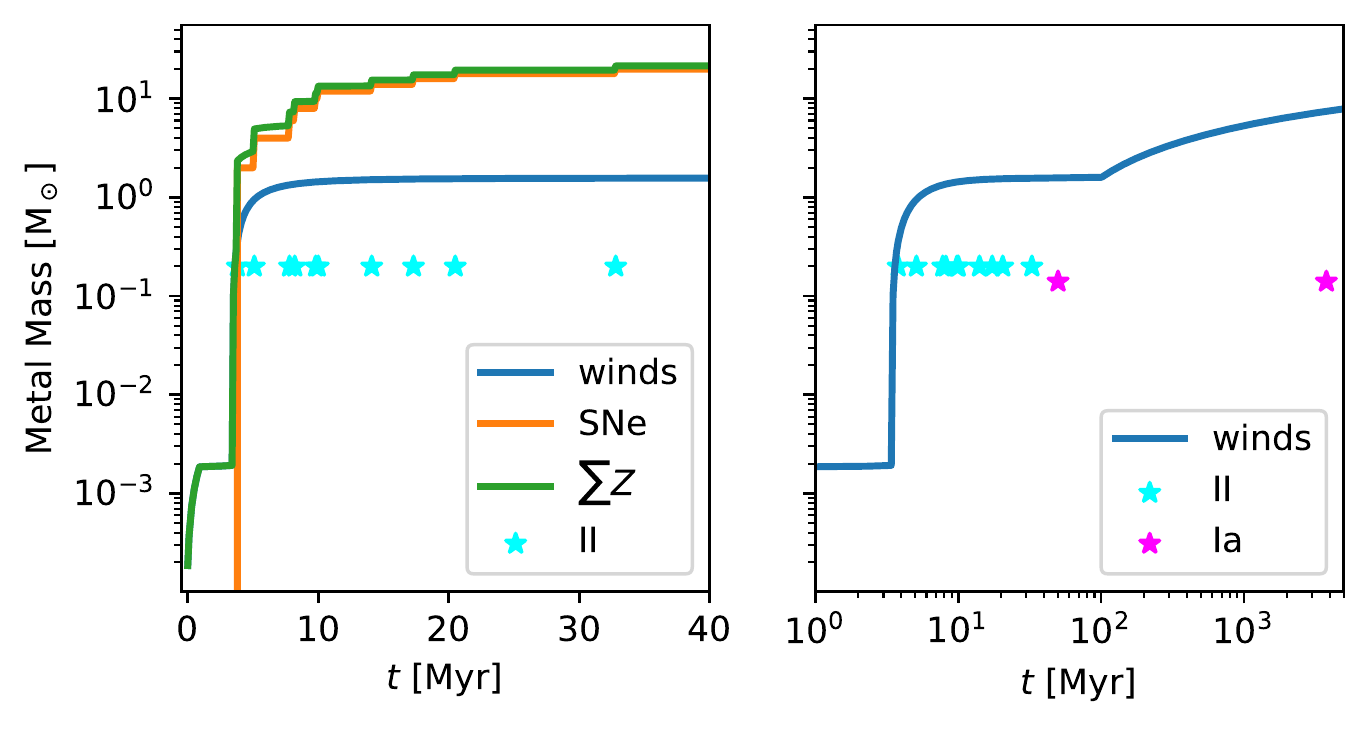}
    \caption{Metal contribution from the evolution of a 1000 \msun~{\tt starss} particle. The left plot shows the 40 Myr evolution of early Type-II SNe era with SN events annotated (cyan stars) and the metal contributions from winds, SNe, and total contributions ($\sum Z$).  Right: the contribution from stellar winds across 5 Byr, with Type-II and Type-Ia events annotated in cyan and magenta, respectively. }
    \label{fig:metal_contribution}
\end{figure}
We present the metal contribution across the lifetime of a \starss particle in Figure \ref{fig:metal_contribution}.  On the right, we show the short-time evolution that includes O-B SNe and their associated winds, with SN events annotated for the first 40 Myr.  The left plot shows the long-time evolution up to 5 Gyr.  It includes late winds from AGB stars starting at 100 Myr, and finite possibility for Type-Ia SNe, with occurrences annotated in magenta stars. At late times, the metal contribution from winds becomes comparable to that from Type-II SNe.

\subsection{Coupling Method}


\begin{figure}
    \centering
    \includegraphics[width=0.46\textwidth]{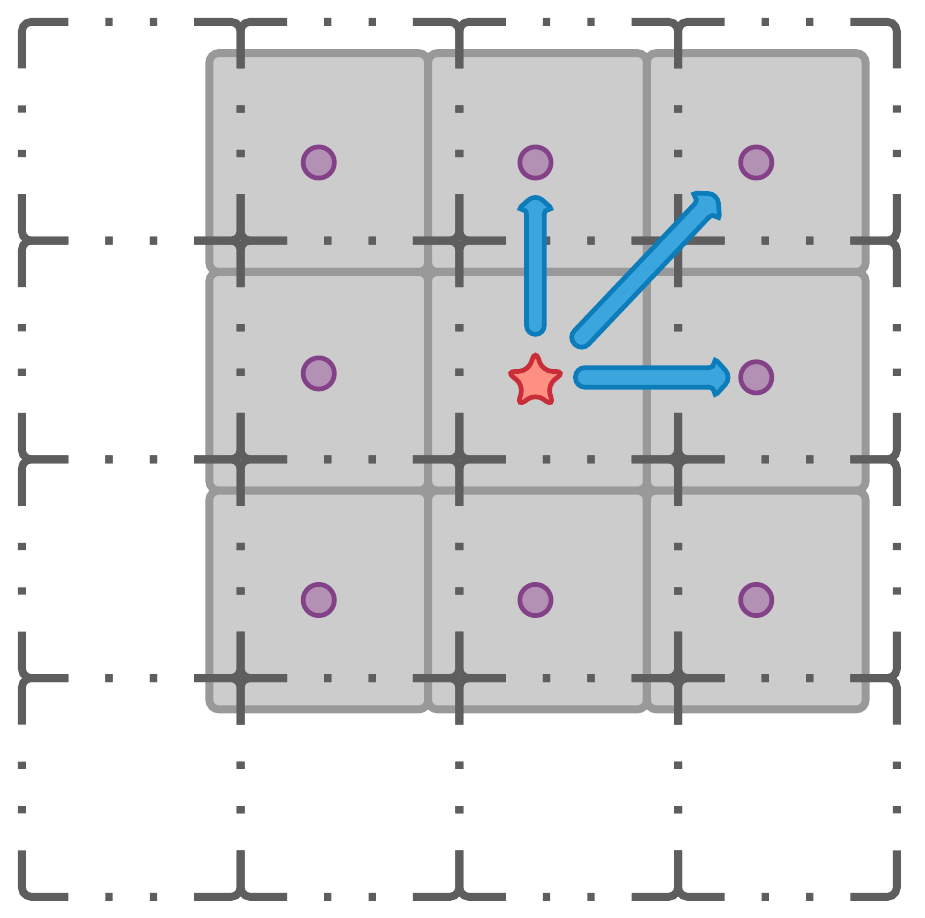}
    \caption{2D example of deposition method.  The feedback source (star) is coupled to neighboring cloud particles (circles) at spacing $dx$ from the source; all feedback quantities are calculated at this radius, with momentum having a vector quantity indicated by blue arrows.  The cloud particles are then cloud-in-cell deposited to the computational grid (hashed black grid). There is no quantity coupled at the source particle, however, the calculated shell mass is removed via CIC from the central cells centered on the source particle.}
    \label{fig:cic_example}
\end{figure}


Given $M_{\rm ej}$, $M_{z,{\rm ej}}$, $E_{\rm ej}$, and $p_{\rm cpl}$ from a source particle, $S_a$, we now describe coupling those quantities to the grid in an isotropic and conservative method that maintains very small linear error.  Shown in the 2D example in Figure \ref{fig:cic_example}, we create a virtual cloud of $3^3-1$ coupling target particles ($S_b$, purple circles) spaced at $dx$ from the feedback source ($S_a$, red star).  This method generates a fixed geometry, and maintains the unique position of $S_a$ within its host cell.  Each $S_b$ receives an equal fraction of energy, mass, metal, and momenta to couple to the grid, so the final quantity coupled at each $S_b$ is 
\begin{equation}
    M_b = \frac{M_{\rm ej}+M_{\rm s}}{26}\\
\end{equation}
\begin{equation}
    M_{z,b} = \frac{M_{z,{\rm ej}} + M_{z,{\rm s}}} {26}
\end{equation}

\begin{equation}
    {\boldsymbol p}_b =  \frac{p_{\rm cpl}}{26} \hat {\boldsymbol r}_{b\rightarrow a}\\
\end{equation}
where $\hat{\boldsymbol r}_{b\rightarrow a}$ is the unit vector from $S_a$ to the $S_b$ particle and the factor of 26 represents the 26 particles in the 3-dimensional virtual coupling cloud.  The above expressions do not include energy: we couple kinetic energy determined by the momenta coupled to the cell as $E_{{\rm k},b} = |{\boldsymbol p}_b|^2/2M_b$.  The thermal energy ($E_t$) coupled is the remainder of the energy budget, i.e., $E_t = E_{\rm ej} - \sum_b E_{{\rm k},b}$.  If $dx > R_{\rm c}$, then the thermal energy is reduced to account for unresolved $P$d$V$ work using $E_t = E_t (dx/R_{\rm c})^{-6.5}$~\citep{thornton1998, hopkins2018}.  Finally, the thermal energy is coupled in the same manner as kinetic with $E_{t,b} = E_t / 26$.  With known quantities for deposition, each $S_b$ virtual particle is coupled to the computational domain via cloud-in-cell deposition.

\section{Idealized Tests}

\label{sec:testing}

To test the \starss supernova feedback algorithm, we used the {\tt TestStarParticle} problem in \enzo.  This problem sets the star particle at the center of the box (plus 1/2 cell width) with uniform density.  To test resolution dependence, we used $\rho = 9.79 \times 10^{-23}\simeq 45$ cm$^{-3}$ in baryon density while varying the cell width, $dx$ between $0.5$ pc $ \leq dx \leq 50$ pc.  The simulation domain was initialized with temperature $T=1000$ K and $Z = -1$ (absolute metallicity 0.001295).  The test utilizes all the standard \enzo physics capabilities: 6 species primordial gas chemical evolution (H I, H II, e$^-$, He I, He II, He III), radiative cooling, and metal-line cooling using 4-D Cloudy lookup tables \citep{smith2009, ferland2017}.  

At time $t = 0.00025$ Myr, a single SN as modeled by \starss is coupled to the grid.  The simulation continues for 1 Myr, and we record the terminal momentum as the maximum measured during that time. Taking the maximum observation is motivated by the behaviour in momentum while varying resolution: in resolved cases where the deposition represents the free-expansion or ST phases (e.g., $dx < 3$ pc) the momentum increases until the terminal value, and decreases afterward, however, in unresolved cases, $\boldsymbol{p}_t$ is directly coupled to the grid.  The initial value of $\boldsymbol{p}$ represents the terminal solution exactly, which decreases after coupling to the mesh due to cooling.  

\begin{figure}
    \centering
    \includegraphics[width=0.48\textwidth]{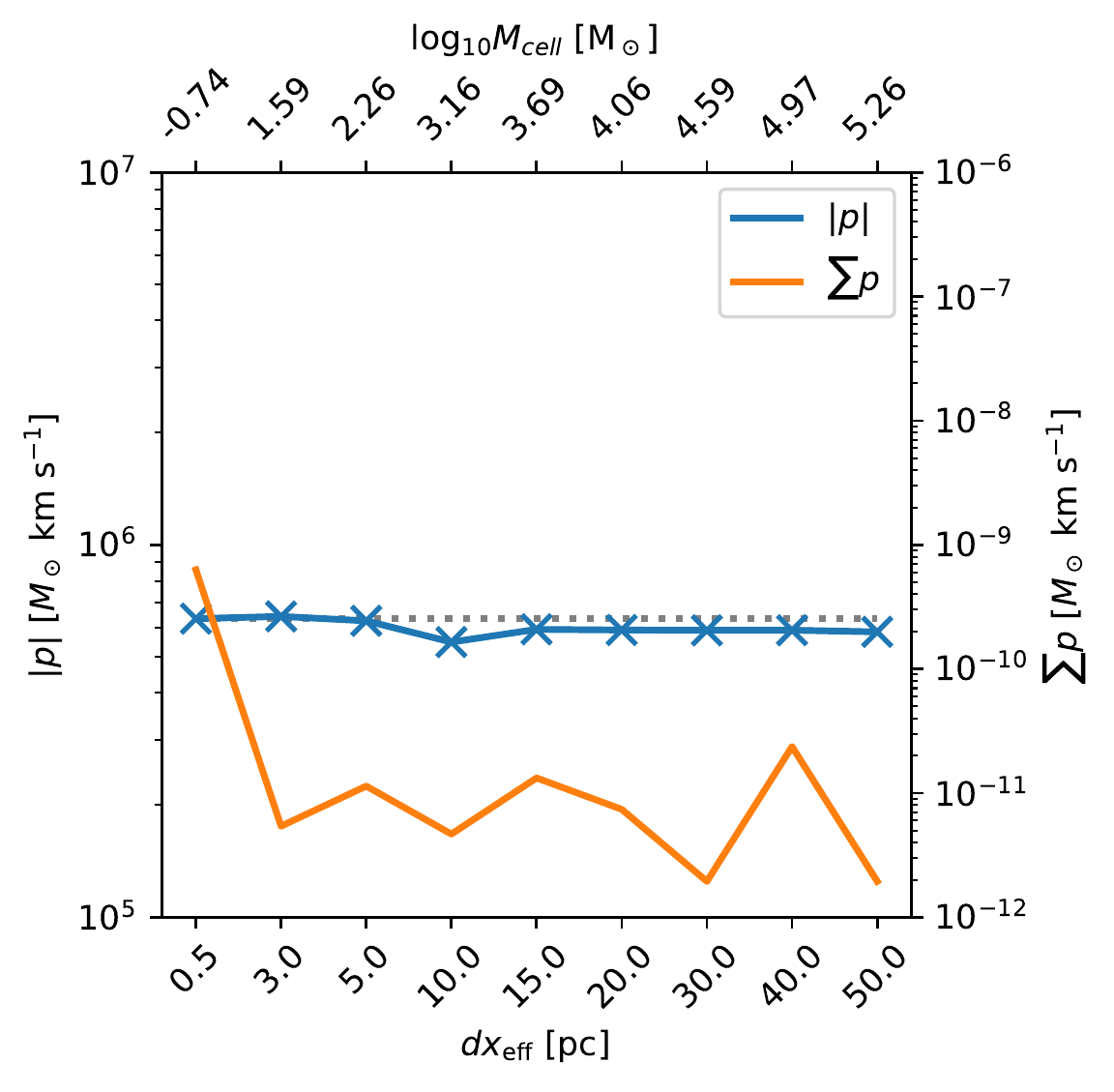}

    \caption{Terminal momentum of ideal tests while varying cell width with fixed $n_b$.  Cell width (dx) annotated on lower axes, corresponding cell gas mass on upper.  Linear momentum error is shown in orange corresponding to the right vertical axis.  Since the density is fixed, we expect the same momentum across all samples, which is the behavior shown in blue.  Expected momentum from the fully-resolved (0.5 pc) test is shown in grey.  It is well matched by \starss across all spatial resolutions tested.}
    \label{fig:res_mom_scale}
\end{figure}

Results from resolution tests are shown in Figure \ref{fig:res_mom_scale}.  The true solution (grey), obtained by a fully resolved simulation with $dx = 0.5$ pc, is closely matched by \starss (blue) in all test cases.  In addition, the linear momentum error ($\sum {\boldsymbol p}$) is $\leq 10^{-9}$ for all test cases. This test does not use the reduction in coupling beyond $R_{\rm fade}$, which results in a sharp drop in coupled momenta for $dx > 30$, finally coupling negligible momenta by $dx = 50$.  

    \begin{figure*}[t]
        \centering
        \includegraphics[width=0.85\textwidth]{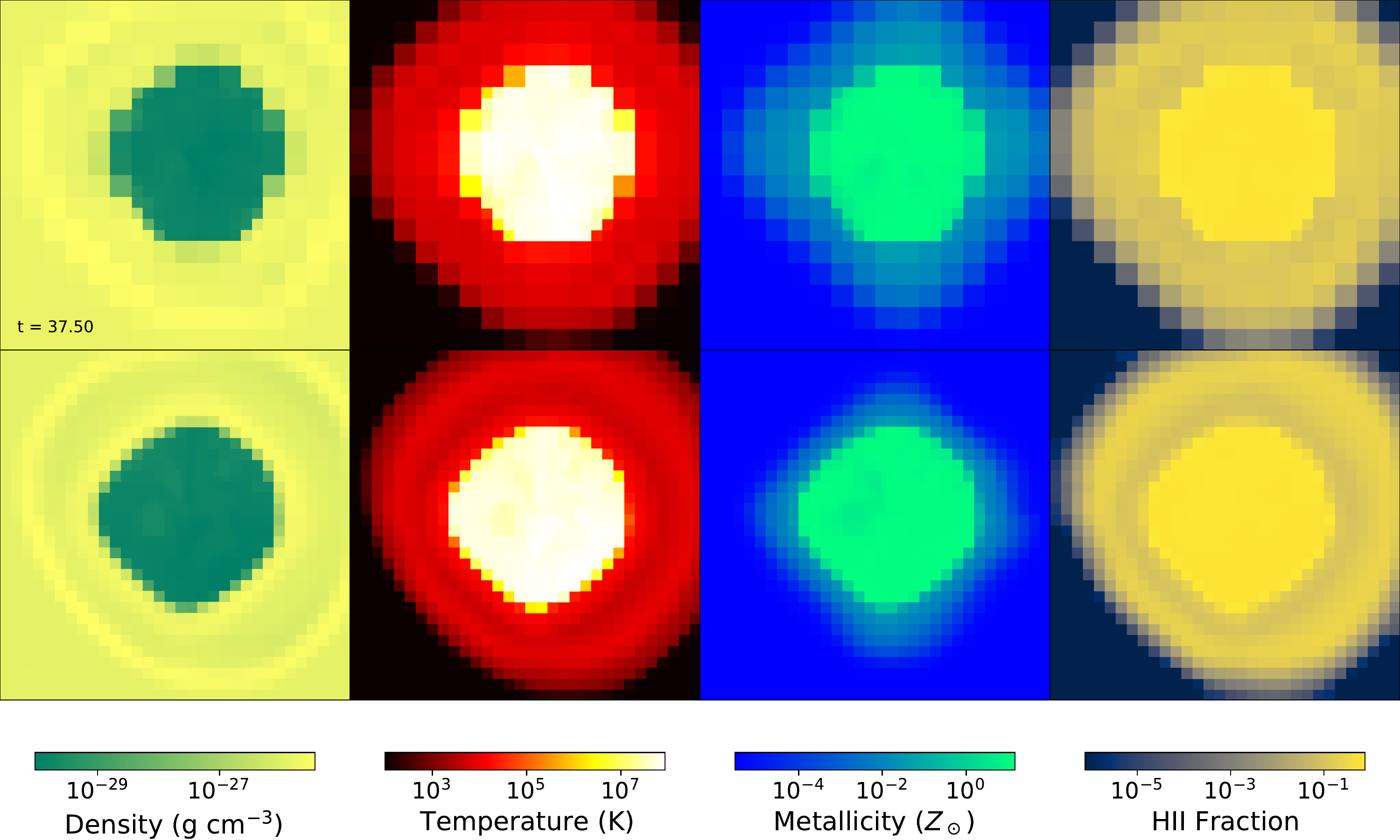} \label{fig:lowres_halo}
        \caption{Ideal halo test comparing two spatial resolutions. Given identical initial density perturbations, we expect very similar results regardless of spatial resolution. Top shows a single particle 20 Myr after creation in a $10^7$ M$_\odot$ halo with a Navarro-Frank-White (NFW) profile \citep{navarro1997} with $16^3$ root-grid and 3 levels of AMR yielding a maximum resolution of 6.25 pc.  Bottom shows the identical initial conditions using a $64^3$ root grid and 3 levels AMR to achieve 1.56 pc maximum resolution.  The final state across these tests is very consistent.  The slight anisotropy is due to the particle being offset from the halo center in the initial conditions.}
        \label{fig:ideal_halo}
    \end{figure*}

    To test the full feedback framework of \starss within a controlled environment, we use an ideal spherical halo.  Specifically, the success of \starss hinges on generating similar results irrespective of spatial resolution, so we performed several tests varying the spatial resolution of a test halo, but holding the baryon and dark matter density constant.  This test uses identical cooling physics as the ideal single supernova test, but now includes radiation from the star particle in the form of ionizing radiation, heating, and photon momentum coupling to the gas as documented in \cite{wise2012b}. The star particle of 1000 \msun is positioned at the center of the halo, slightly offset from the center of the box with width 800 pc.  In the first low resolution test (LR), the domain has $16^3$ root-grid cells and 3 levels of AMR on dark matter density so that the center of the halo has $dx = 6.25$ pc.  The second high resolution test (HR) has $64^3$ root-grid cells with 3 levels of AMR producing a maximum resolution of $1.56$ pc. The refinement criteria of each simulation are tuned such that the center of the halo is at the maximum AMR level. The background density of the box is set to $7.0\times 10^{-26}$ g cm$^{-3}$ giving the center of the halo number density $n_b = 2.2$ cm$^{-3}$ at $t=0$ in both cases.  The simulation allows for the full probabilistic feedback of \starss, and proceeds for the duration where Type-II SNe have finite probability, 37.53 Myr. Although the initial central number density of the halo has $n_b = 2.2$ cm$^{-3}$, early radiation pressure reduces the density surrounding the star particle to $n_b \sim 0.8$ cm$^{-3}$ in both LR and HR. Since the number of SN events is not deterministic, the simulations were repeated until we observed a similar number of events during the lifetime of the particle.  Figure \ref{fig:ideal_halo} shows slices through the simulation domain at the final output, 37.50 Myr. We observed 16 events in the LR (top) and 15 in the HR.  Both show qualitatively similar behavior, including in the magnitude of density, temperature, and metal density within the superbubble remnant (SBR).  At the final output, we measure the momentum in the ``shell" of the SBR and find that $|{\boldsymbol p}| = 7.12\times 10^6 \textrm{ and } 7.42\times10^6$ M$_\odot$ km s$^{-1}$ for the HR and LR respectively, representing a $\sim 4.2\%$ difference.  The magnitude of difference in momenta, $\sim 3\times 10^5$ \msun km s$^{-1}$, is very similar to the expected momenta from a single terminal-phase SNR.

\section{Cosmological Simulation Tests}
\label{sec:cosmo_starss}
We now explore the performance of \starss within the framework of cosmological simulations.   We prefer the multiple galaxy-halos of cosmological simulations so that we can compare \starss across multiple star forming halos and multiple halo mass ranges within a single simulation.  Each simulation uses identical cosmological parameters with with $\Omega_m = 0.3111$, $\Omega_b = 0.048975$, $\Omega_k = 0$, $\Omega_\lambda = 0.6889$, $H_0 = 0.6766$, $\sigma_8 = 0.811$, $n = 0.965$~\citep{planck2014}.  For easy comparison, each uses identical initial conditions generated using {\tt MUSIC} \citep{hahn2011} on a 256$^3$ root-grid with (2.61 Mpc)$^3$ volume.  With these parameters defined, the dark matter particle mass is $2.34\times10^4$ \msun and the average initial baryon mass per cell is $1.17\times10^3$ \msun.  For refinement criteria, we consider baryon and dark matter overdensity, where the cell is refined if $\rho_x/\bar\rho_x > 3$, for the baryon or dark matter density $\rho_x$, and $\bar\rho_x$ refers to the simulation averaged quantity.  In addition, we use an exponential factor to enforce super-lagrangian refinement: the the mass within a cell to cause refinement is given by $M_x \geq 3 M_{i} \times 2^{-0.6l}$ for level $l$ and $M_i$ as the initial average baryon or DM mass per cell in the simulation.

We use identical physical and chemical models across all \starss simulations: 9-species primordial gas chemistry including H, H$^+$, He, He$^+$, He$^{++}$, e$^-$, H$_2$, H$_2^+$, H$_2^-$; 4-dimensional metal line cooling considering density, metallicity, electron fraction, and temperature as determined by Cloudy lookup tables \citep{smith2009}; radiative feedback including photon momentum coupling to the gas using the Moray ray tracing solver \citep{wise2011} with \starss particles as sources; finally, we include a redshift dependent Lyman-Werner H$_2$ dissociating radiation background to model sources from outside the simulation region (\cite{xu2016}, Eq. 1).


\begin{figure*}
    \centering
    \includegraphics[width=0.95\textwidth]{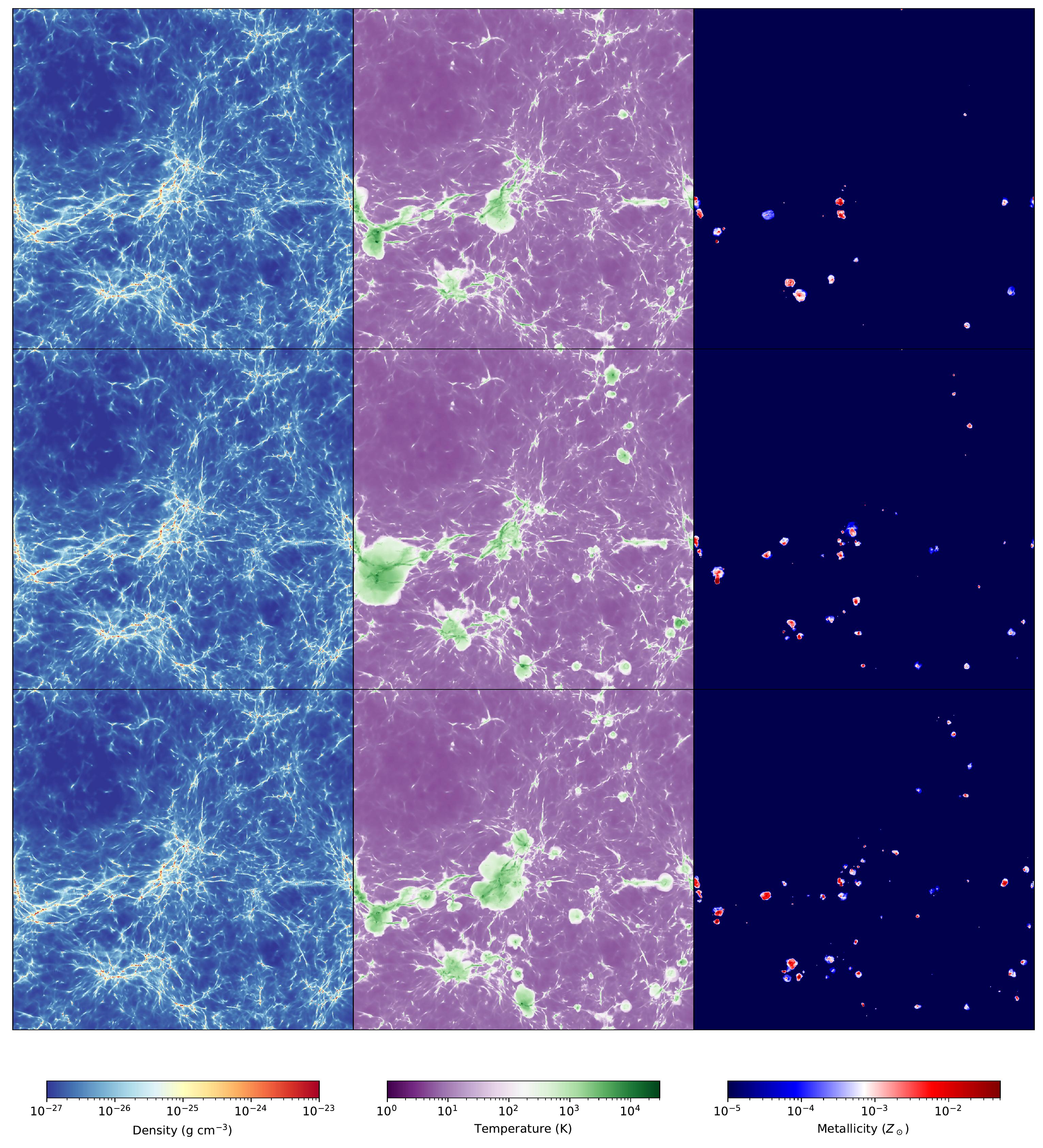}
    \caption{Comparing simulations with identical mass resolution varying maximum spatial resolutions at $z=13.91$.  The top, middle, and bottom show the 4L, 5L and 6L simulations respectively.  Note that since there is a probabilistic element in both the star formation and supernova routines, we do not expect perfect agreement among the rows.}
    \label{fig:sts_sidebyside}
\end{figure*}
The cosmological simulations used to validate \starss are shown as projections in Figure \ref{fig:sts_sidebyside}, and with parameters summarized in Table \ref{tab:sts}. We vary the maximum number of levels of refinement (denoted by the integer in the run name) while keeping the mass resolution constant as well as all other simulation parameters. This has the effect of varying the finest cell-width, which at $z=15$ varies from 39 pc to 9.75 pc. Qualitatively, the simulations produce very similar results, however minor differences, e.g., in the size of bubbles in metallicity, are noticeable.  We can see that the volume is largely unenriched by $z=13.91$, and by using temperature as a proxy for ionization state, most of the volume remains neutral (in reference to H, assuming ionized H gas would have $T\gtrsim 10^4$ K).  While the simulations are comparable in the metallicity field, signifying consistent SN feedback across resolutions, the temperature field shows systematic differences, particularly between 4L and 5L simulations, where 4L has noticeably smaller high-temperature regions that are fewer in number.  These systemic differences are less pronounced when comparing the 5L and 6L simulations, indicating that the star formation algorithm is more consistent at those resolutions.  The remainder of this section is dedicated to quantitative comparisons across these three resolutions.

\begin{table}[]
    \centering
    \caption{Summary of \starss simulations.}
\centering    \begin{tabular}{|l|c|c|c|c|}
    \hline
        ID & $L_{\rm Max}$ & $Z_{\rm f}$  & $dx_{0}$ & $dx_{15}$ \\\hline\hline
        4L  & 4 & -5.5             & 624 pc       & 39 pc    \\
        5L  & 5 & -5.5             & 312 pc       & 19.5 pc      \\
        6L  & 6 & -5.5             & 156 pc       & 9.75 pc      \\\hline\hline
        4LZ  & 4 & -3             & 624 pc       & 39 pc    \\
        5LZ  & 5 & -3             & 312 pc       & 19.5 pc      \\
        6LZ  & 6 & -3             & 156 pc       & 9.75 pc      \\\hline\hline

    \end{tabular}\vspace{3pt}\\
    {Note: The differences between simulations is shown here; the metallicity floor ($Z_{\rm f}$) and maximum AMR level varies, while all other parameters are identical throughout.  The finest cell-width is shown for $z=0$ ($dx_0$) and $z=15$ ($dx_{15}$).}
    \label{tab:sts}
\end{table}


\begin{figure}[t]
    \centering
    \subfloat[$Z_{\rm f}=-5.5$]{
        \includegraphics[width=0.4\textwidth]{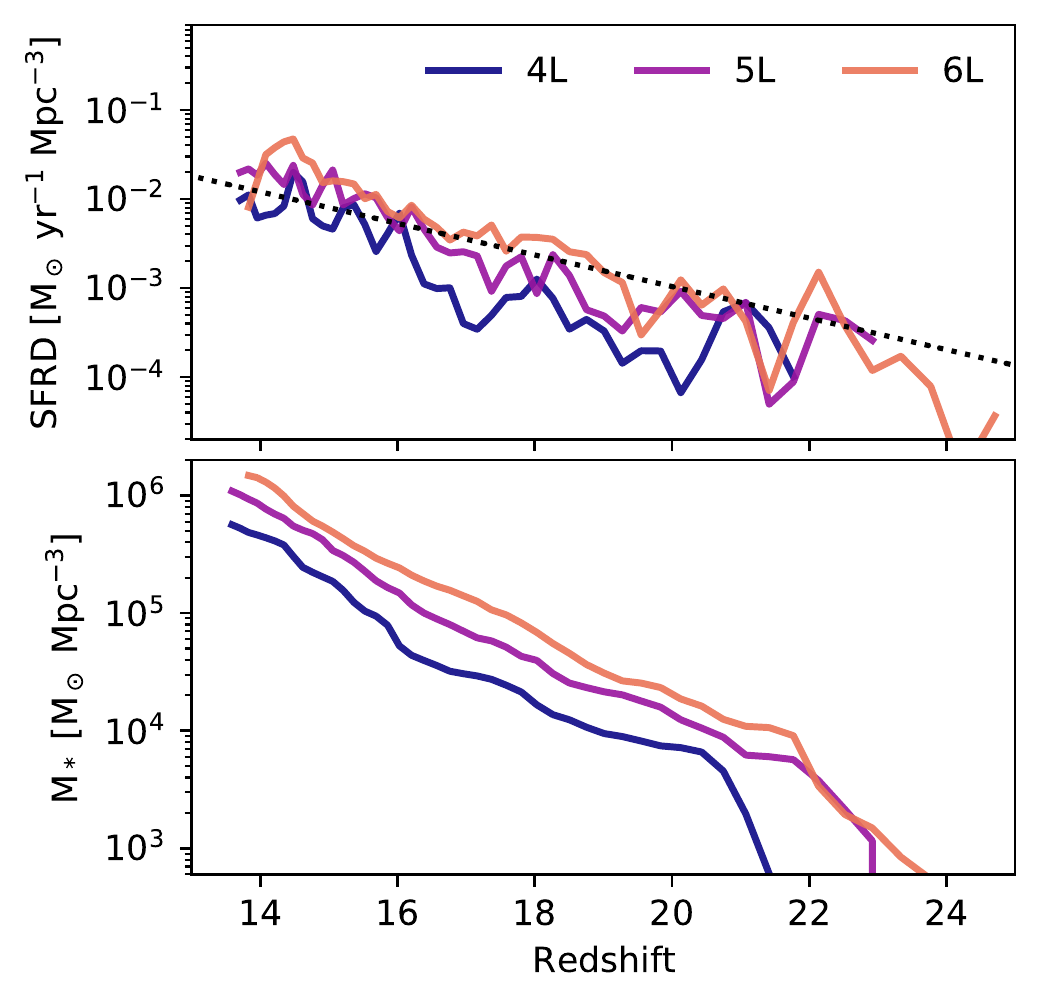}
        \label{fig:sfrd_lz}}
    \subfloat[$Z_{\rm f}=-3$]{
        \includegraphics[width=0.4\textwidth]{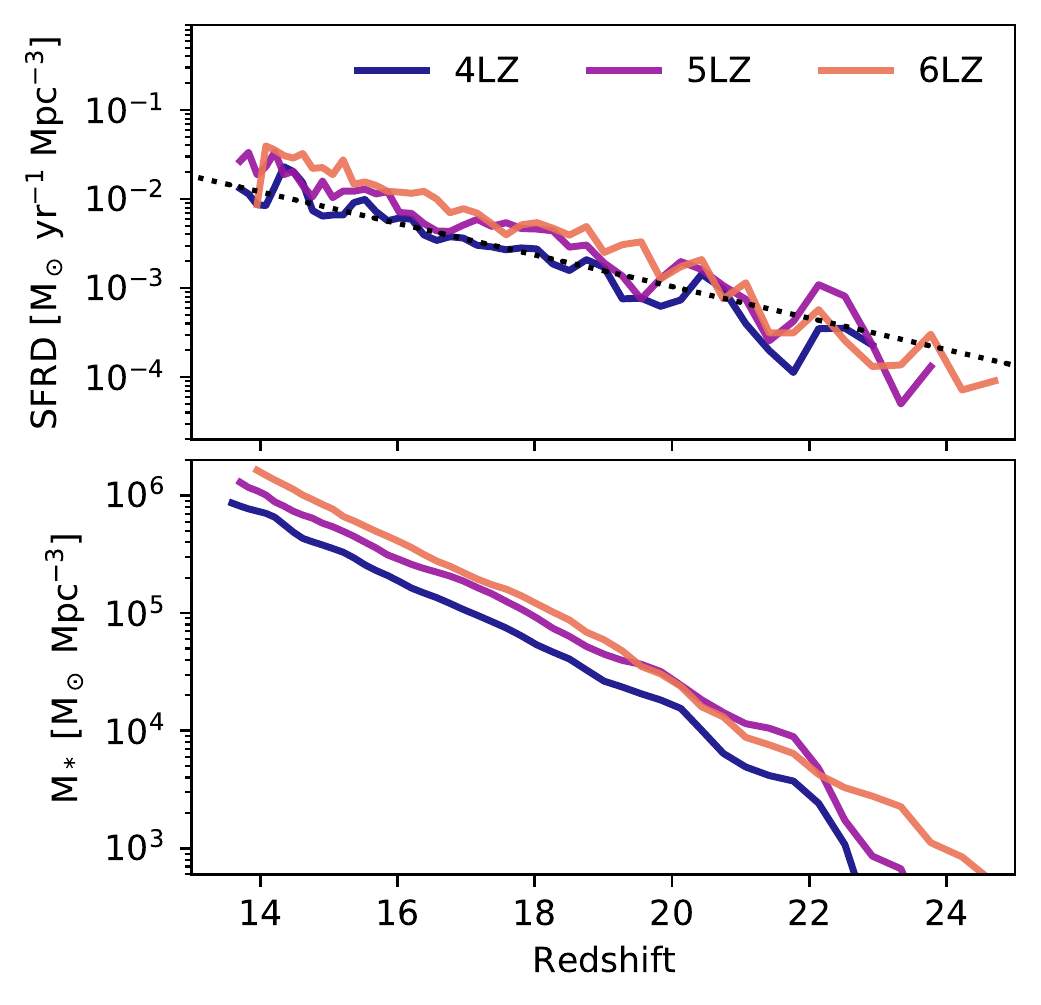}
        \label{fig:sfrd_hz}}

    \caption{SFR density and formed stellar mass density for varying resolution using \starss.  All resolutions use the same star formation and feedback parameters. Panel \ref{fig:sfrd_lz} uses a metallicity floor of $Z_{\rm f}=-5.5$, while \ref{fig:sfrd_hz} uses $Z_{\rm f}=-3$.  5LZ and 6LZ simulations converge to nearly identical behavior with the higher metallicity floor, but have slight differences in stellar mass with the low metallicity floor. The black dotted line is an approximate fit to the SFRDs with SFRD = $3.65\exp(-z/2.45)$.}
    \label{fig:sfrd_comps}
\end{figure}

We can quantify the convergence of the \starss algorithm by examining the cosmic star formation rate (SFR) and stellar mass formed from a global perspective that includes both initial conditions for a given resolution.  In Figure \ref{fig:sfrd_comps}, we present the SFR density (SFRD) and stellar mass density  for all \starss simulations.  Figure \ref{fig:sfrd_lz} uses the same metallicity floor ($Z_{\rm f}$) as the critical metallicity of the \textit{Phoenix Simulations} with $Z_{\rm f} = -5.5$.  While 5L and 6L have very similar cumulative stellar mass, 4L lags by $\sim -0.3$ dex.  However, 4L maintains similar slope in cumulative mass, suggesting that the effect of resolution is to delay star formation, but not to change the rate of star formation after it has begun.  Figure \ref{fig:sfrd_hz} shows the same set of simulations, but using a metallicity floor more common in modern work of $Z_{\rm f} = -3$.  The higher metallicity floor enables more efficient collapse of star forming regions due to enhanced cooling, which has acted to negate most resolution effects seen in Figure \ref{fig:sfrd_lz}, but has little effect on the resolved 6L case.  With $Z_{\rm f} = -3$, 5LZ and 6LZ have converged to nearly identical behavior, and the stellar mass deficit in 4LZ is reduced to only $\sim 0.2$ dex below 5LZ and 6LZ. The behavior in Figure \ref{fig:sfrd_hz} is our primary motivation for using 5LZ in our comparisons of Section \ref{sec:starnet_comps}. 

\begin{figure*}
    \centering
    \includegraphics[width=0.85\textwidth]{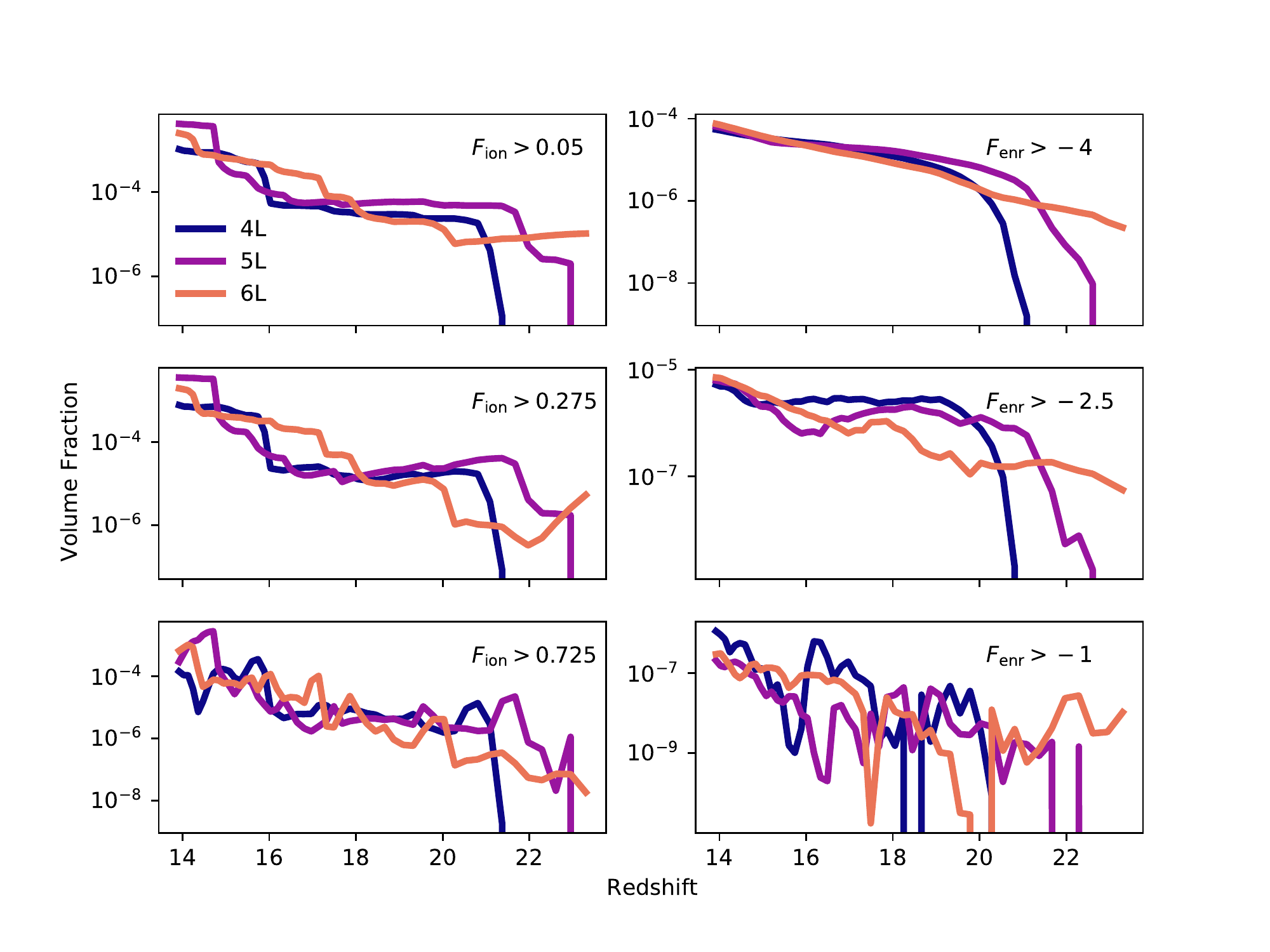}
    \caption{Volume fraction ionized to various levels (Left column), or enriched to varying levels (Right column) by redshift using the second initial conditions set for all resolutions.}
    \label{fig:enr-ion-fractions}
\end{figure*}
To describe how star forming regions interact with the environment from a global perspective, we show the volume fractions ionized and enriched to varying levels in Figure \ref{fig:enr-ion-fractions}.  There is much better agreement in the enriched volume fraction than ionized volume fractions.
Lower resolution (4L, 5L) tend to enrich more of the volume at early times, but the difference between resolutions is largely negligible by $z\sim14$.  The ionized volume fraction still shows deviations at the lowest redshift, and shows large steps corresponding to star formation beginning in new regions.  Exploration to lower redshift would be beneficial to determine whether the differences in ionized volume fraction are reduced or exacerbated by further evolution.

\begin{figure*}

    \centering
    \includegraphics[width=0.95\textwidth]{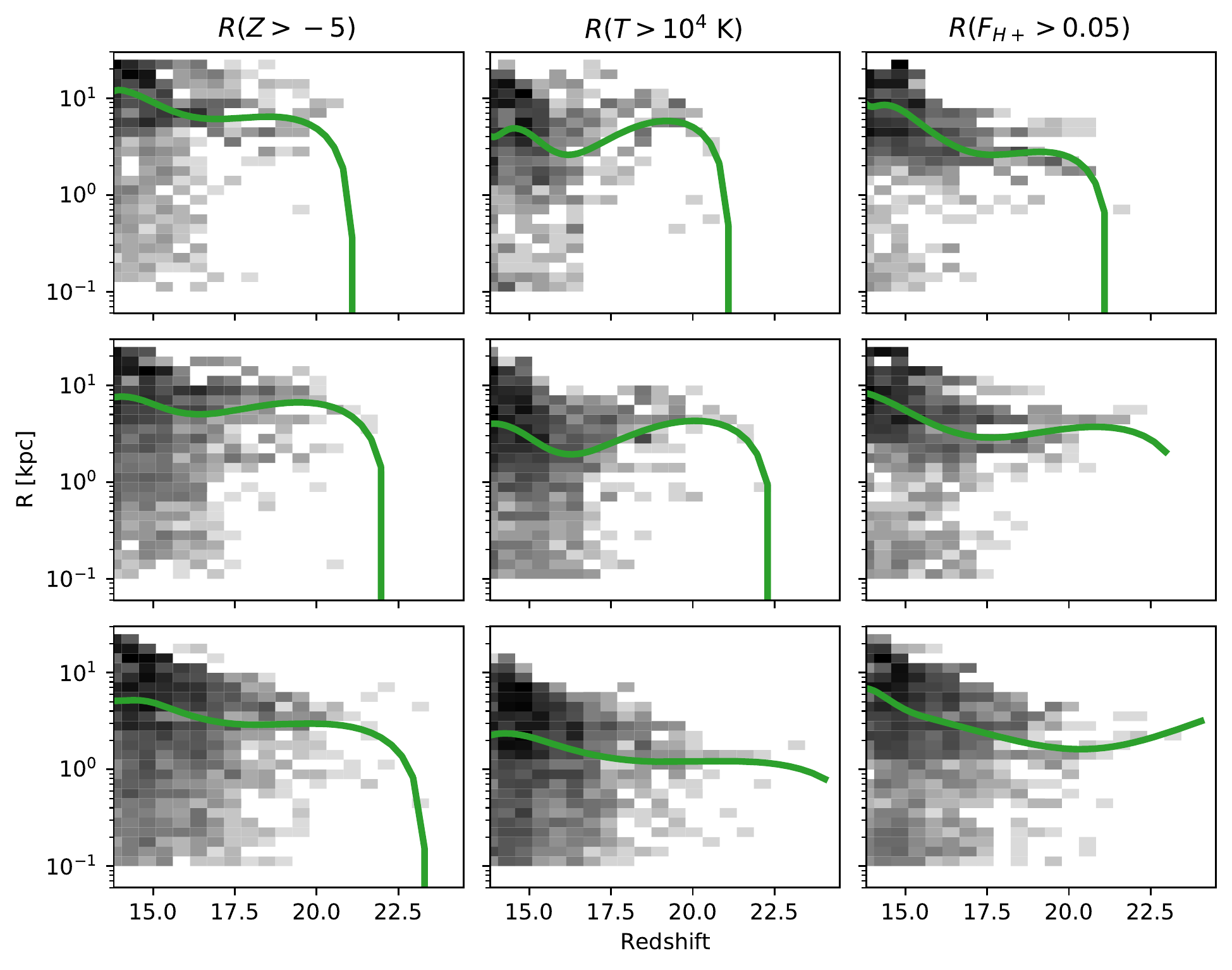}
    \caption{ Comparing the size of feedback remnants across 4L (top), 5L (middle), and 6L (bottom) simulations. The background histogram logs each star forming region, while the plotted line shows a bicubic spline fit to the 50$^{\rm th}$ quantile of remnants in each redshift bin.}
    \label{fig:sts-sizeof}
\end{figure*}

In Figure \ref{fig:sts-sizeof}, we show a histogram of feedback region radius by redshift for all \starss simulations.  We additionally show the mean radius plotted over the histogram.  To obtain this data, we iterate each data output and search for halos with finite stellar mass. If found, we center a 0.1 kpc sphere on the center of stellar mass and record the volume-weighted mean metallicity and HII fraction.  If $\langle Z\rangle > -5$ or $f_{\rm HII} > 0.05$, the sphere is expanded by 0.1 kpc and the averaging is repeated until $\langle Z\rangle < -5$ and $f_{\rm HII} < 0.05$.  Figure \ref{fig:sts-sizeof} clearly shows that there are fewer small feedback regions as resolution decreases, implying that lower-resolution runs struggle to model low stellar mass or young systems.  The ionization fraction is consistent across the presented resolutions, however we note larger metal-rich remnants as resolution decreases. While noticable between 5L and 6L, the effect is much more pronounced in comparing 4L and 6L.  This is likely due to 4L creating fewer but more clustered stellar particles instead of many small, scattered particles.  Those few clustered particles deposit more SNe into the same region. In a positive feedback loop, the higher number of SNe are then deposited into less dense and hotter gas, resulting in higher-velocity expansion and larger super-bubble remnants.  Despite this complication, the quantile fit to the size of remnants in 4L is still $\lesssim 2\times$ that of 6L, despite having 4$\times$ worse spatial resolution.  Examining the background histogram suggests that all resolutions have similar distributions at high radii, and the significant change is in the modeling of younger or less massive stellar systems with smaller radii.
\begin{figure}
    \centering
    \includegraphics[width=0.48\textwidth]{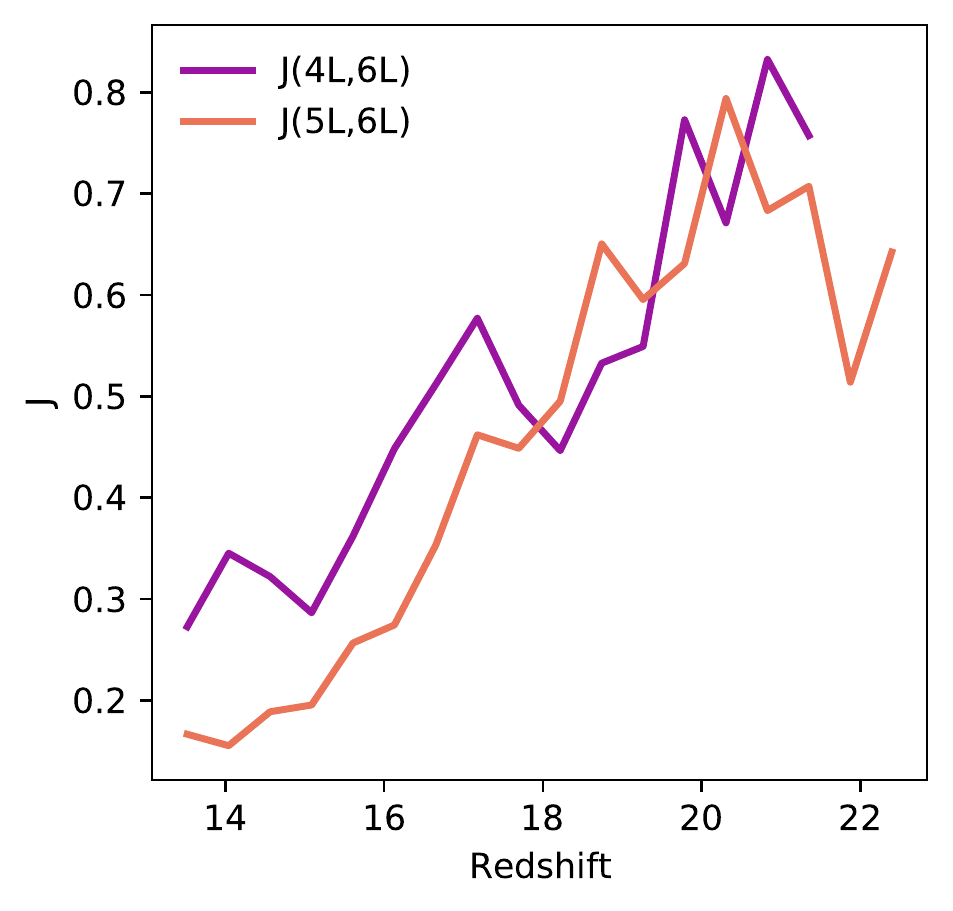}
    \caption{Comparing redshift bins of metallicity radius in Figure \ref{fig:sts-sizeof} by examining Jensen-Shannon distance.  Lower distance implies more closely related PDFs of radii.}
    \label{fig:size_JSD}
\end{figure}

The distribution of remnant regions in Figure \ref{fig:sts-sizeof} can be quantitatively compared by considering each redshift bin as an independent distribution of remnant sizes.  We can then compare each bin using the Jensen-Shannon distance as a metric to compare two probability distributions.  With the probability distribution of radii from simulation A as $P_a$, and from simulation B as $P_b$, we can define the Kullback-Leibler divergence $D(P_a|P_b) = P_a\log (P_a/P_b)$ to compare the PDFs using the Jensen-Shannon distance given by
    \begin{equation}
        J(P_a, P_{b}) = \sqrt{\frac{D(P_a|\bar P_{ab}) + D(P_{b}|\bar P_{ab})}{2}},
    \end{equation}
where $P_{ab}$ is the pointwise mean of $P_a$ and $P_b$.  For this description of $J$, $J=0$ implies identical PDFs, while $J=1$ would indicate completely unrelated PDFs.  We compare the 4L and 5L simulations to the 6L simulation using this method in Figure \ref{fig:size_JSD}.  In general, $J$ is lower when comparing 5L and 6L, and the agreement between the three resolution increases as redshift decreases.  Although 5L appears to be a qualitatively better match in Figure \ref{fig:sts-sizeof}, this metric solidifies that observation.  

\begin{figure*}
    \centering
    \subfloat[$Z_{\rm f}=-5.5$]{
    \includegraphics[width=0.45\textwidth]{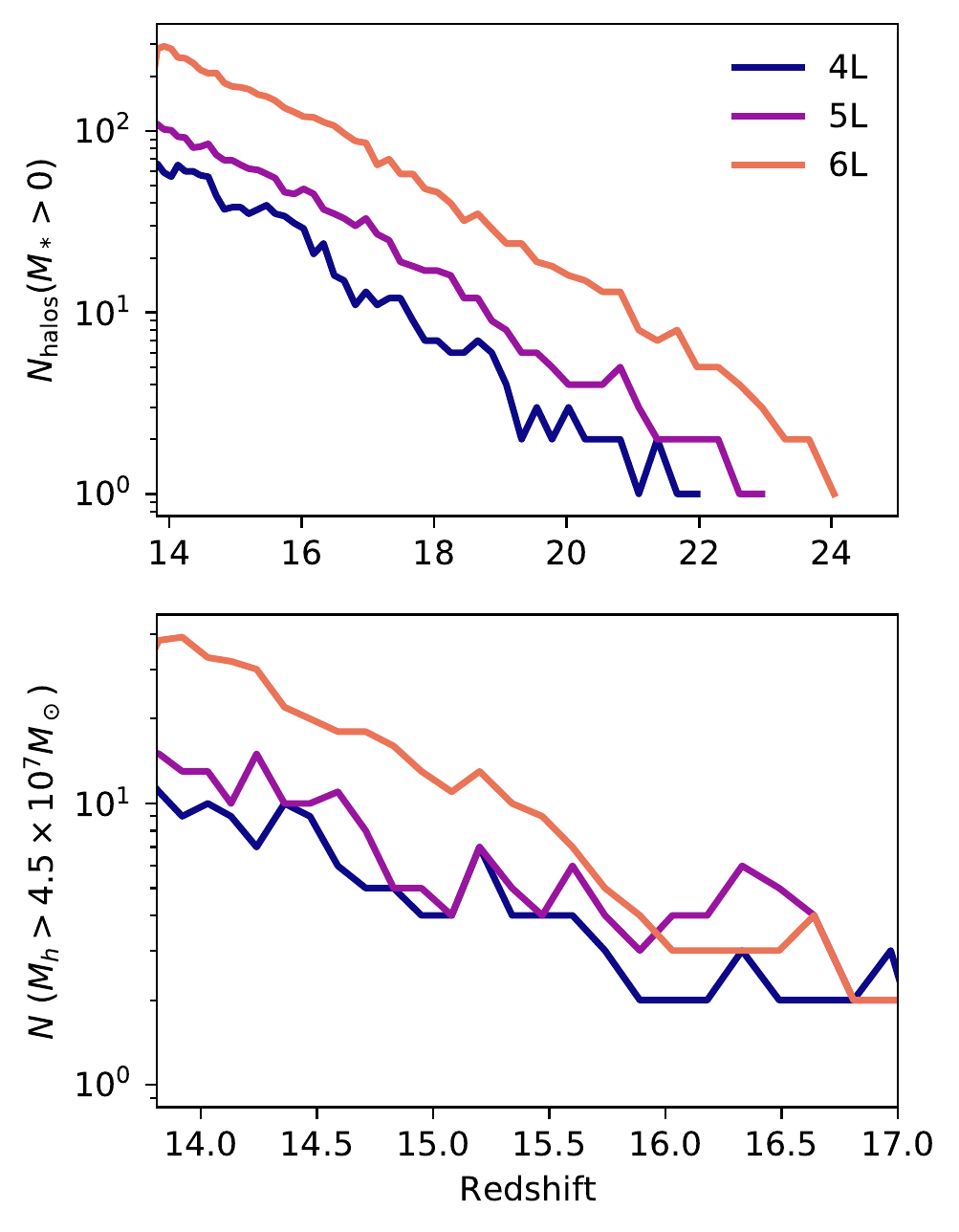}
    \label{fig:halo_count_lz}
    }\hspace{10pt}
    \subfloat[$Z_{\rm f} = -3$]{
    \includegraphics[width=0.45\textwidth]{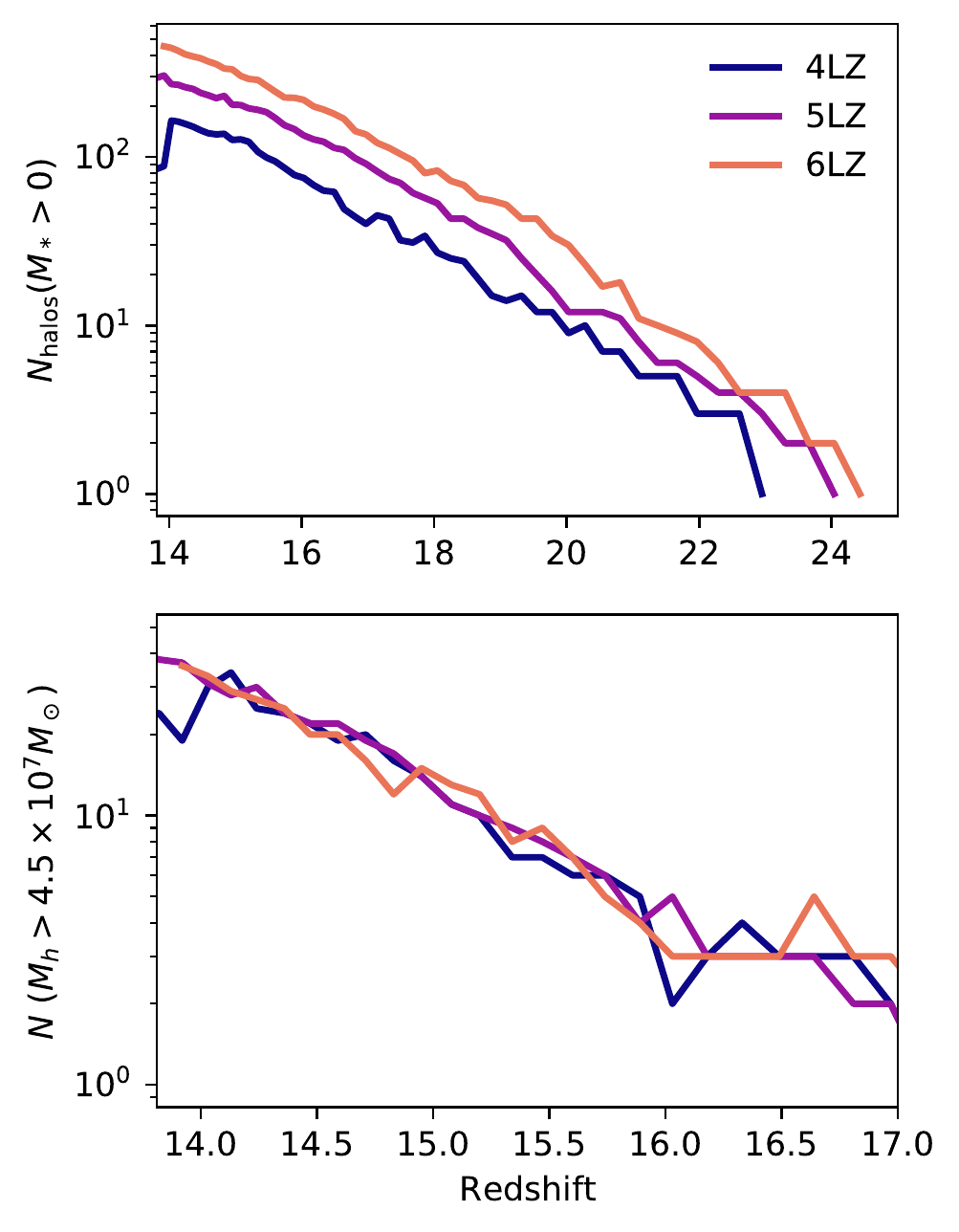}
    \label{fig:halo_count_hz}
    }
    \caption{Star forming halo counts including all IC variations. Top: Number of star forming halos by redshift.  Bottom: Number of star forming halos with $M_{h} > 4.5\times10^7$ \msun by redshift.  Note that the plots do not share range in redshift.  \ref{fig:halo_count_lz} shows the low $Z_{\rm f}$ simulations, while \ref{fig:halo_count_hz} shows $Z_{\rm f} = -3$.  Using $Z_{\rm f} = -3$ has negated much of the discrepancy noted in the $Z_{\rm f} =-5.5$ case.  Factor of $\sim 2$ drops in counts at $z\sim 14$ (4LZ) and $z\sim 15.5$ (6LZ) are due to different end redshifts of simulations within that resolution suite.}  
    \label{fig:halo_count}
\end{figure*}

The figures presented thus far indicate that while the feedback from \starss is largely consistent across resolution scale, the star formation algorithm is not.  This is obvious in Figure \ref{fig:halo_count}, where we present the number of star-forming halos (top) and number of massive $M_h > 4.5\times10^7$ \msun star-forming halos (bottom) in each tested resolution for both $Z_{\rm f} = -5.5$ and $Z_{\rm f} = -3$.  The effect of resolution is to host star formation in fewer halos, particularly for low $Z_{\rm f}$ (Panel \ref{fig:halo_count_lz}). The effect is reduced as the halo mass of interest is increased, however, even at  $M_{h} \sim 4.5\times 10^7$, 6L shows enhanced star forming halo counts.  The effect is also greatly reduced when using $Z_{\rm f}=-3$ (Panel \ref{fig:halo_count_hz}).  While the difference between 5LZ and 6LZ is still $\sim 0.1$ dex for all halos, their behavior is converged for larger halos with $M_{\rm vir} \geq 4.5\times 10^7$ \msun.  In \starss current paradigm of predicting star formation, this relationship between resolution and star-forming halo counts is likely unavoidable. Since force resolution in \enzo decreases with spatial resolution, the simulation cannot resolve peaks and troughs of the hydrodynamic fields as accurately. This leads to lower magnitudes of density peaks reducing the ability of the gas to self-shield, as well as reduced accuracy in modeling gas flows, limiting star formation by our $\nabla\cdot\boldsymbol{v} < 0$ star formation criterion.

\section{\starnet: Surrogate Models of Primordial Star Formation and Feedback}
\label{sec:starnet}

Our primary goal with this work is to generate heterogeneous metallicity initial conditions that resemble the effects of \pas, but with modification so that the spatial resolution of the simulation can be relaxed.  The prior section introduced the \starss algorithm to address the resolution-dependence of metal-enriched star formation; in this section we present the new method of generating metallicity initial conditions in cosmological simulations. Where prior methods would set a ``metallicity floor," where the metallicity everywhere in the volume takes on a set value, or ignore metallicity effects on the first and second generations of stars, here we will detail our method to generate an heterogeneous metallicity field by considering where \pas will exist and modelling their subsequent metal injection without the extreme resolution requirements of prior works.  Using the inline Python analysis capability in \enzo, we incorporate deep learning models to predict \pa formation (\starfind (W21)) and regression models of \pa region effects (W22) into a single framework (\starnet) that can evaluate simulations {\it in situ} for primordial star formation sites and deposit a rudimentary approximation of their effects.  Although every effort was made to streamline and optimize the inline Python for \starnet, the data structure of \enzo combined with implementation limitations on the inline Python means that these runs were limited in the following ways: each deposition of a \pa remnant requires access to all levels of the computational grid; therefore, we cannot utilize load balancing algorithms that would separate child grids from their parents. Additionally, there is no communication across tasks, so we must reduce the possibility of remnant regions that cross root-grid boundaries (each root-grid tile is a work-unit managed by one MPI task in \enzo).  We accomplish this by using the minimum number of tasks that can support a 256$^3$ root-grid simulation.  Because of these limitations, this work only explores the very high-redshift regime of $z\sim 15$: future work to integrate the StarNetRuntime into \enzoe will be able to fully explore lower-redshift simulations that probe the epoch of reionization, $z\sim 6$.  Accordingly, this work is a ``proof-of-concept'' and first attempt at exploring the application of DL methods to active simulations.  The primary goals of this work are to elucidate whether the \piii era is strictly necessary for modelling, to determine the usability of these DL models,  and to inform future implementations and efforts in similar regards.
    
    \subsection{\starnet}
    \begin{figure*}
        \centering
        \includegraphics[width=0.95\textwidth]{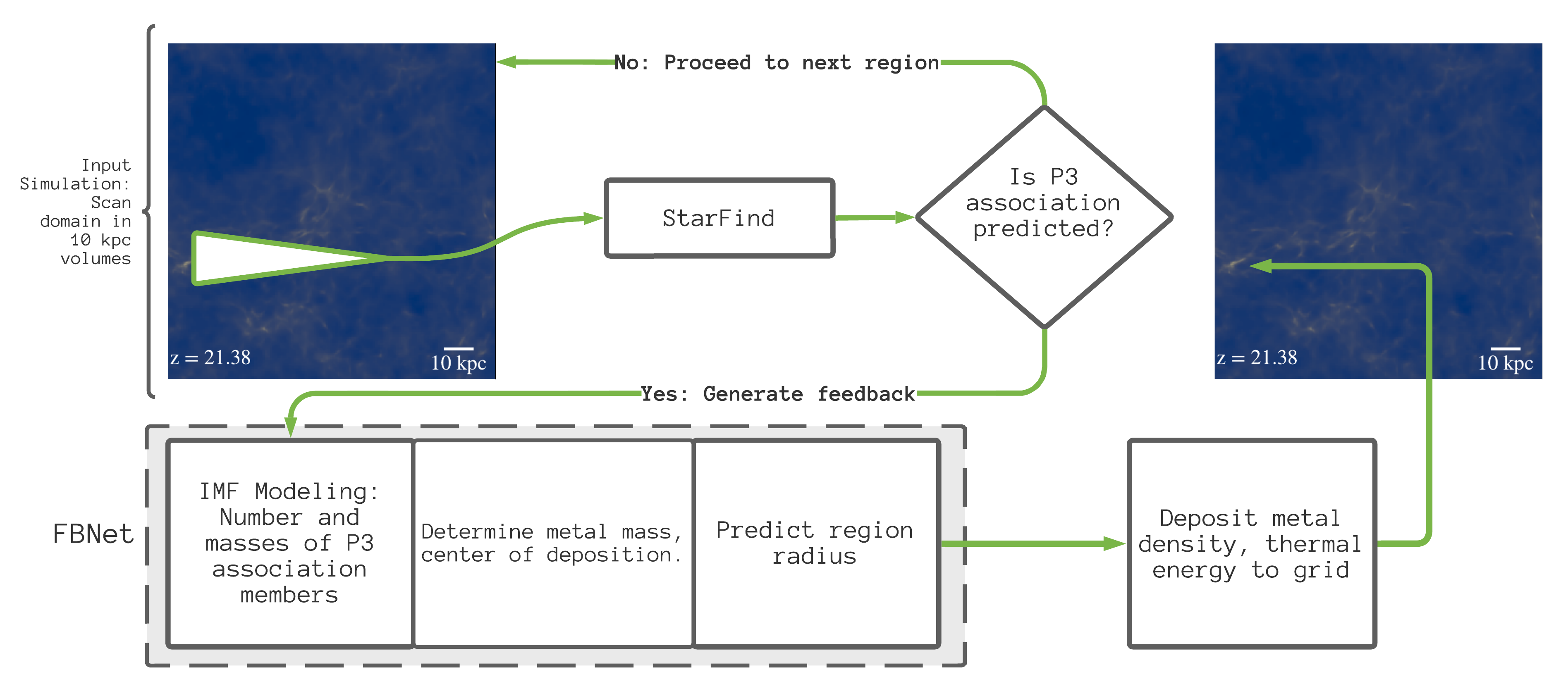}
        \caption{Simplified work-flow diagram of \starnet.}
        \label{fig:starnet-workflow}
    \end{figure*}
    \label{fig:starnet}
    A simplified work-flow of \starnet is presented in Figure \ref{fig:starnet-workflow}. Every 5 Myr, the simulation is paused and the active computational domain is sent to \starnet for evaluation.  First, each grid with AMR level $l=2$ and mean density above $3\times$ the cosmic mean density is tiled into (10 comoving kpc)$^3$ volumes.  Each volume needs to represent AMR level 6, despite the fact that each volume may include data from various levels and is assuredly not uniformly the same resolution.  To bring data from $l=2$ to $l=6$, we copy simulation data at $l=2$ to an inference grid, $G$, use trilinear interpolation to achieve $l=3$, and then copy the simulation data at level 3 to $G$.  This process is repeated at increasing levels until the desired resolution, where $G$ has 64$^3$ dimension. $G$ is then passed into trained models for \starfind to perform inference. 
    
    If \starfind predicts a \pa, we begin calculating a simplified feedback solution that represents the final evolved state of the \pa (from a metallicity perspective). \starfind predicts a group of voxels in $G$ that are participating in star formation, where we assume that the center of the region is at the center-of-mass of those positive prediction voxels.  We use the statistical relations described in W22 to determine the number and masses of \piii stars within the association and use the mass-SN yield parameterization of \enzo to derive the metal mass originating from the \pa, $M_{\rm III}$.   We then use the linear regression models of W22 to predict the radius of the feedback region, $R_{\rm P3}$. From the center of the predicted \pa, the computational grid within $R_{\rm P3}$ is modified using the following: 
    \begin{itemize}
        \item Primordial metal density, $\rho_{\rm ZIII} = M_{\rm III} / (4\pi R_{\rm P3}^3 / 3)$\footnote{All spherical-volume based calculations are corrected at deposition to account for depositing into cubical grid cells}.
        \item Temperature, $T=10^4$ K.
        
    \end{itemize}
    We make no effort to model the changes in ionized species, or the abundances of primordial species--all chemical species retain the same fraction of density that existed before to maintain mass conservation. In effect, it is assumed that the metals are uniformly mixed with the pre-existing primordial gas.  While these modifications are very crude, they satisfy several requirements: the metallicity field is not homogeneous on large scales (outside the feedback region) and the high temperature ionizes H gas and will reduce the H$_2$ fraction within the region.  In addition, since $R_{\rm P3}$ and $\rho_{\rm ZIII}$ are calculated from prior simulations statistics, each ``bubble" has a broad range of possible metallicities, which prevents each star forming region from attaining the exact same metallicity floor before \pii star formation commences.
    
    \begin{figure*}
        \centering
        \includegraphics[width=0.8\textwidth]{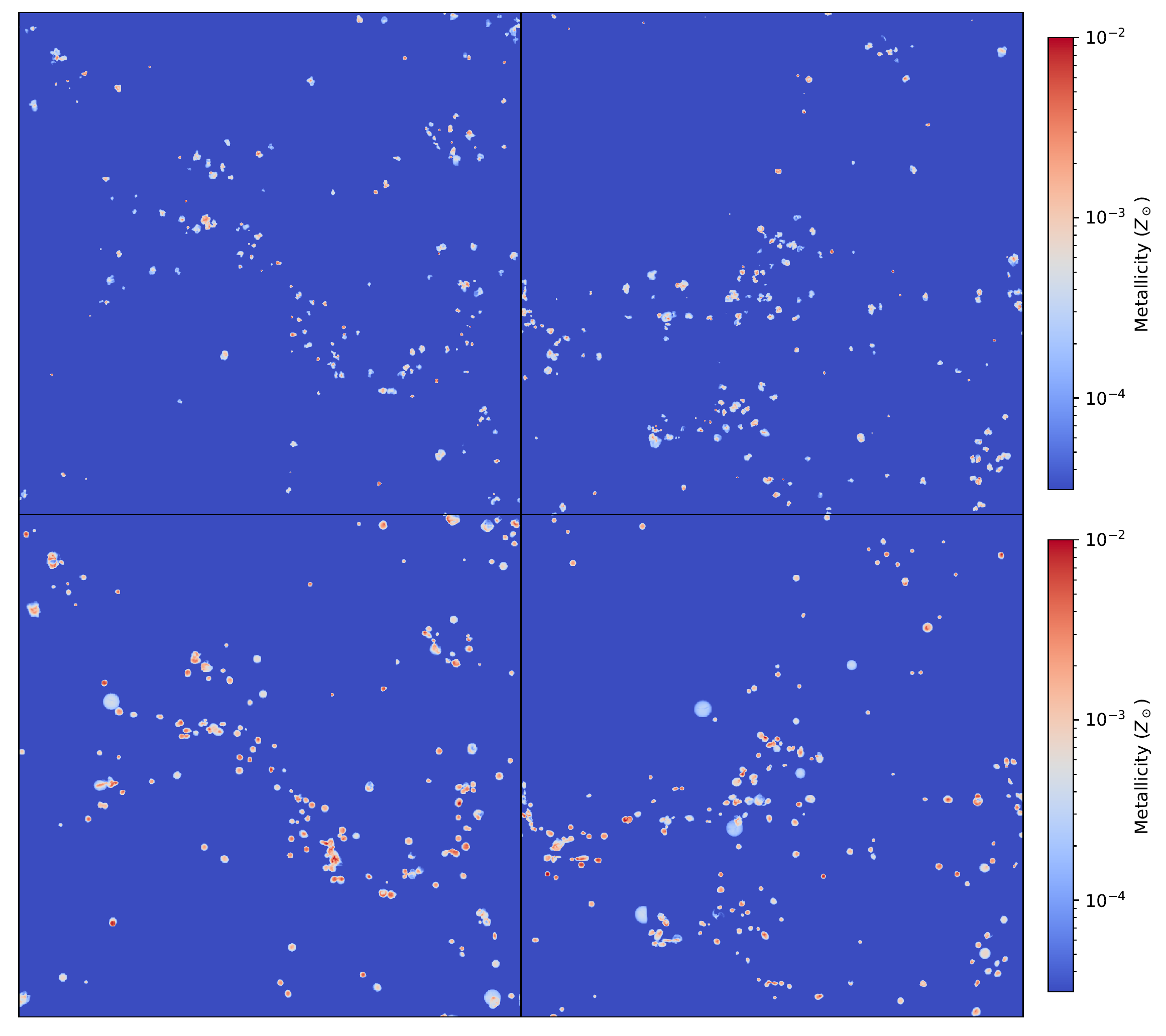}
        \caption{Top: Metallicity field sourced from \piii stars within the PHX256-1 and PHX256-2 simulations (W22).  Bottom: Metallicity field as predicted by \starnet.  Both rows show the state at $z=14.95$.}
        \label{fig:sn-phx-comp}
    \end{figure*}
    \subsection{Validating \starnet}
    
     \starnet is composed of the results of previous projects which have never been combined into a single module until this work.  To validate \starnet, this section shows a brief comparison between \starnet models and their ground-truth, the \textit{Phoenix Simulations} (PHX).   A visual projection of the primordial metallicity field is shown in Figure \ref{fig:sn-phx-comp} comparing the metallicity field in the PHX256-1 and PHX256-2 simulations (top) to the predicted regions found by using \starnet in a simulation matching the initial conditions of the PHX, but with no \pii star formation enabled.  There is obvious agreement between many regions in the projections, however there are some notable differences.  Some regions have been predicted early by \starnet, within the lower plot.  As well, it appears that the regions are, generally, larger in \starnet.  This is an artifact of \starnet's design: it predicts a ``final state" for the feedback region, estimating the impact of primordial stars 16 Myr after the first star forms.  Since the regions in the PHX suite are at varying stages of evolution, there are many regions that have just formed in PHX, but have the final state predicted by \starnet.  It is notable however, that the magnitude of metallicity  seems to agree well between the two methods as well. 
    
    \begin{figure}
        \centering
        \includegraphics[width=0.48\textwidth]{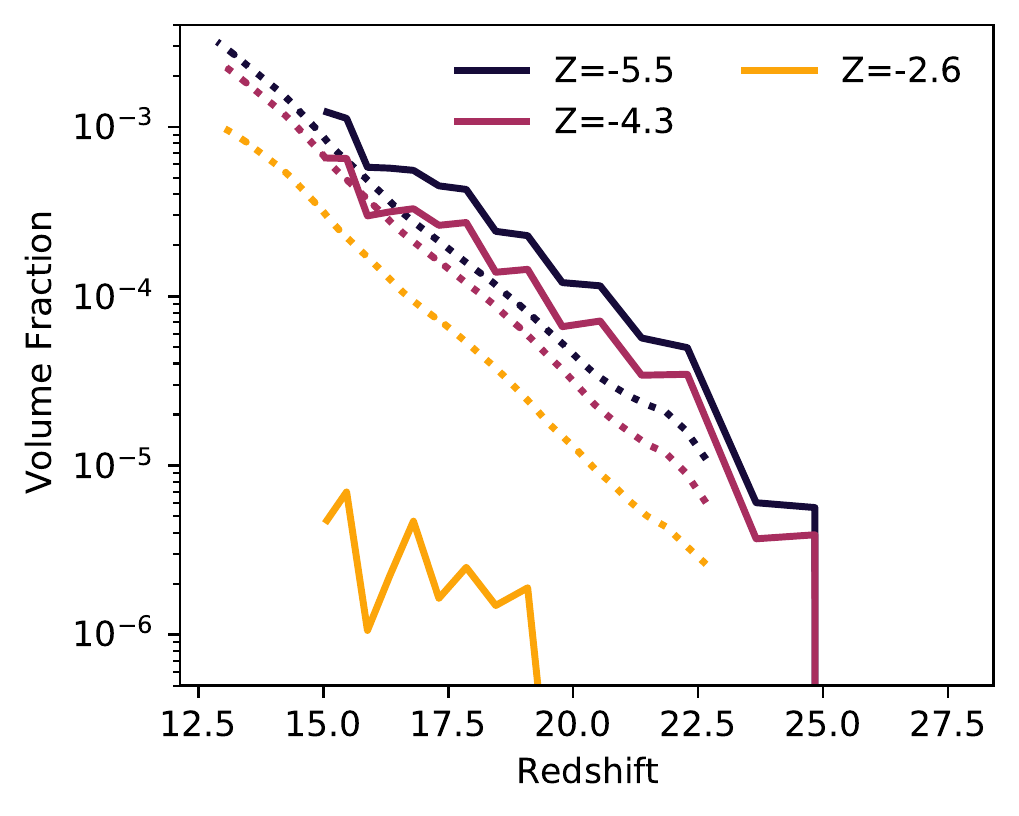}
        \caption{Volume of simulation enriched above the given metallicity, as generated by \starnet without including \starss star formation and feedback.  The hashed lines show the corresponding fraction as measured in the {\em Phoenix Simulations'} PHX512~(W22).  Since there is no on-going star formation in the \starnet simulation, comparisons between ionized volume fractions are uninformative.}
        \label{fig:inline_enr_frac}
    \end{figure}


  To further validate that \starnet is producing expected size of feedback region and in a reasonable number of events, we compare the fraction of volume enriched to varying degrees in Figure \ref{fig:inline_enr_frac}.  The volume fraction enriched to low levels, i.e., $Z=-4.3, -5.5$, is tracked by \starnet to $\sim0.3$ dex.  Notably, at high metallicity, \starnet predicts a much lower enriched fraction than was found in the PHX512.  Many of these can be attributed to \starnet depositing the ``final state'': there is no point where \starnet models the early phases of expansion--the SBRs existence is binary--so the early phase of SBR is ``fast-forwarded" to the final state that would exist 16 Myr later.  This effect causes the enriched volume to be incremented in steps instead of a smooth transition between times, which leads to temporary larger errors.  Algorithmic effects aside, the existence of these relatively high-$Z$ regions will enable a broad range of metallicity for second-generation star formation, further increasing the diversity of protogalaxies enabled by \starnet.    

    \begin{figure}
        \centering
        \includegraphics[width=0.48\textwidth]{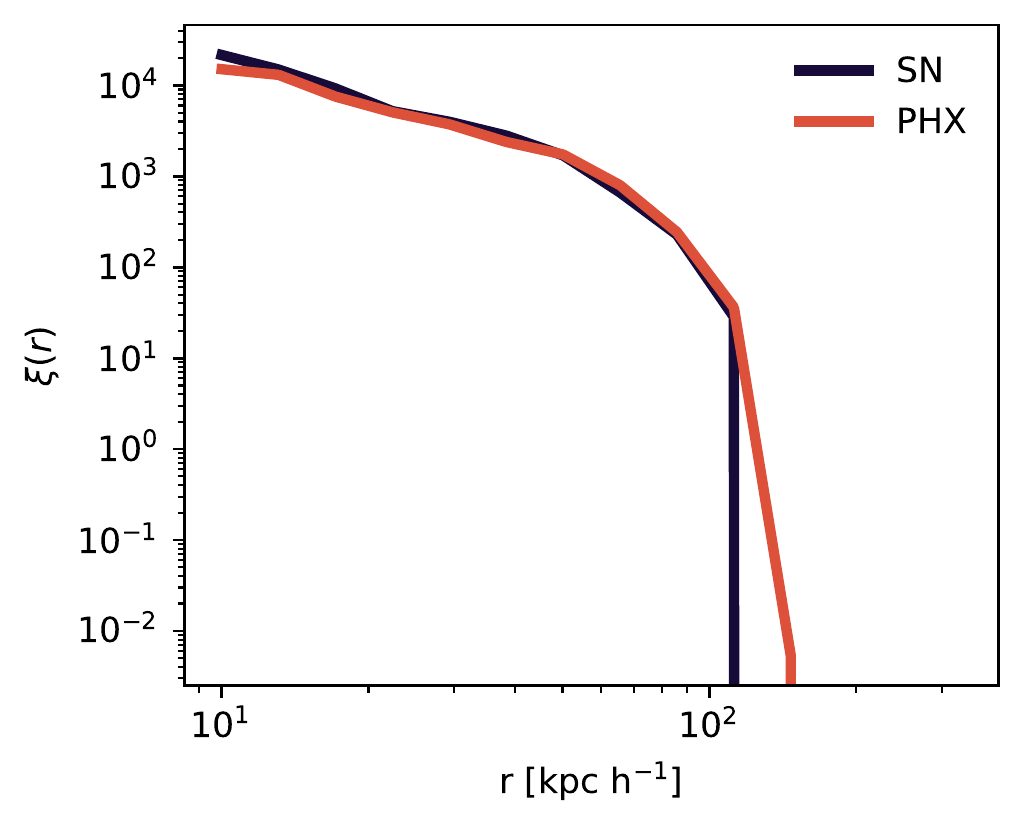}
        \caption{Two point correlations of protogalaxies enriched with primordial gas in the SNET and PHX256-1 simulations.  With minor deviation at low-$r$, $\xi(r)$ is consistent across simulations.}
        \label{fig:tpcf-sn-phx}
    \end{figure}
    
    To further discuss the distribution of enrichment by \starnet, we compare the two-point correlation functions ($\xi(r)$) for halos enriched by Pop III SNe both in PHX256-1 and cosmological simulations using \starnet for primordial enrichment, where both versions share identical initial conditions. We calculate $\xi(r)$ as 
    \begin{equation}
        \xi(r) = DD(r)/RR(r)-1
        \label{eqn:xi}
    \end{equation}
    where $DD(r)$ is the number of halo pairs with separation equal to $r$, $RR(r)$ is the number of randomly distributed pairs that would have separation $r$, with radii are split into bins of width $\delta r$\footnote{{\tt halotools}:  https://halotools.readthedocs.io}.
    Ideally, \starnet would duplicate $\xi(r)$ from PHX256-1, indicating that \starnet had enriched an identical distribution of halos as was present in the PHX.  Indeed, Figure \ref{fig:tpcf-sn-phx} shows that sn5l-1 and PHX256-1 have nearly identical $\xi(r)$, with only minor deviation at high and low $r$. At least some of the low-$r$ deviation can be attributed to the final-state modeling of \starnet--small regions that would have expanded and merged together are modeled as a single merged region.

\section{Measuring the Impact of Heterogeneous Metallicity Initial Conditions from \piii Stars}
\label{sec:starnet_comps}
    In order to quantify the effect of pre-enrichment by \piii stars, we perform five simulations:  
    
    \begin{enumerate}
        \item sn5L-1: \starnet simulation that requires $Z\geq -5.5$ prior to \starss particle formation.
        \item sn5L-2: same as sn5L with different initial conditions.
        \item sn5l-noZ: \starnet simulation with no metallicity prerequisite--\starss particles can form regardless of gas metallicity in the host cell, however \starnet still generates a metallicity field due to \piii star formation.
        \item sts5l-1: \starss simulation from Section \ref{sec:cosmo_starss}; this has no \piii treatment, instead initializing the metallicity field at $Z=-3$ and requiring $Z\geq -3$ for \starss particles to form.  This modified metallicity floor is the 5LZ simlulation set noted in Section \ref{sec:cosmo_starss}, and is more representative of typical metallicity floors used in, i.e., FIRE-2.
        \item sts5l-2: same as sts5l with changed initial conditions matching that of sn5l-2.
    \end{enumerate}
\newcommand{\sn}{SNET\xspace}
\newcommand{\sts}{STS\xspace}
    
    Broadly, these simulations fit two categories:  dependent on or independent from initial enrichment. Below, the terms \textit{\sn} or \textit{\sts} shall refer to the combined statistics of sn5l-1,2 or sts5l-1,2 respectively.  Our primary comparison is between the \sn and \sts suites to identify the difference between simulations using metallicity floors and those requiring pre-enrichment from \piii SFF.   \sn uses a very low critical metallicity compared to the metallicity floor of \sts.  Our motivation for doing so is because the training data to create \starfind and \starnet were generated from the \textit{Phoenix Simulations} suite where \pii star formation can only proceed in gas enriched to $Z_c\geq -5.5$.  Modifying the critical metallicity floor of \sn would require further testing to ensure that \starnet can enrich gas appropriately to higher levels without further training. Hence, in this work, we compare the two as methodological approaches, but not as an attempt to compare similar critical metallicities or floors.  The sn5L-noZ simulation provides a unique dataset.  Since each \starss particle inherits its metallicity from the host cell where it formed, this simulation will identify stars that formed in metal-free or low-$Z$ gas by their metallicity fraction, while still allowing for star formation to proceed in the pre-enriched regions created by \starnet.  This simulation will, in particular, provide the fraction of stars in a final galaxy that would not have formed if we enforced a metallicity requirement for star formation.


\subsection{Global Simulation Statistics}
\begin{figure}
    \centering
    \includegraphics[width=0.48\textwidth]{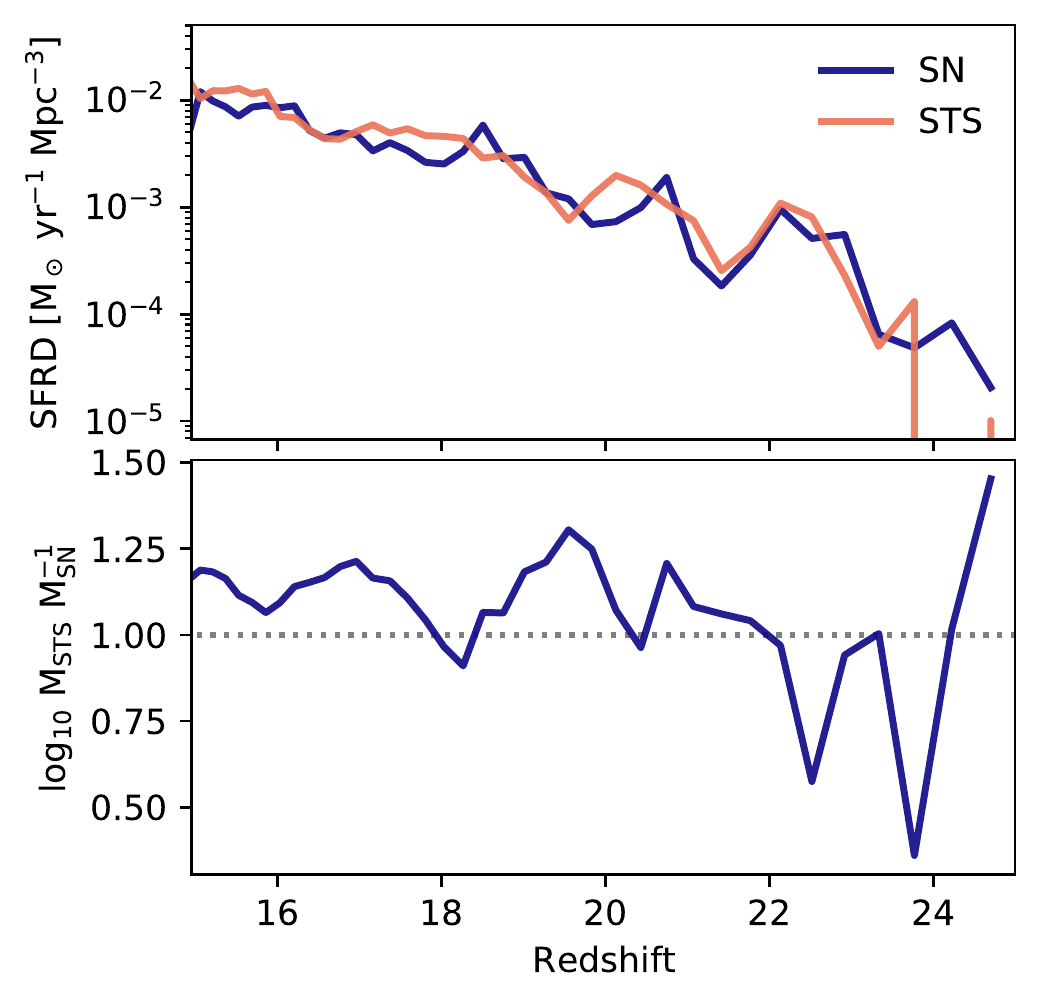}
    \caption{Comparing SFRD of \sn and \sts. The lower panel shows the error in stellar mass obtained by using a uniform metal density IC as in the \sts.}
    \label{fig:sn-sts-sfrd}
\end{figure}
    In Figure \ref{fig:sn-sts-sfrd}, we show both SFRD and stellar mass difference across \sn and \sts with stellar mass diference given by $M_{*, {\rm STS}} /M_{*, \rm{\sn}}$.  \sn and \sts have comparable SFRD at all times (top), however the small differences in SFRD accumulate to rather large errors in cumulative stellar mass formed (bottom). At very early times ($z > 21$), \sts drastically under-produces stars, having as much as 50\% less stellar mass.  This early trend is reversed at later times, where \sts overproduces stars and eventually has $\sim20\%$ excess stellar mass compared to \sn.
    
\begin{figure}
    \centering
    \includegraphics[width=0.48\textwidth]{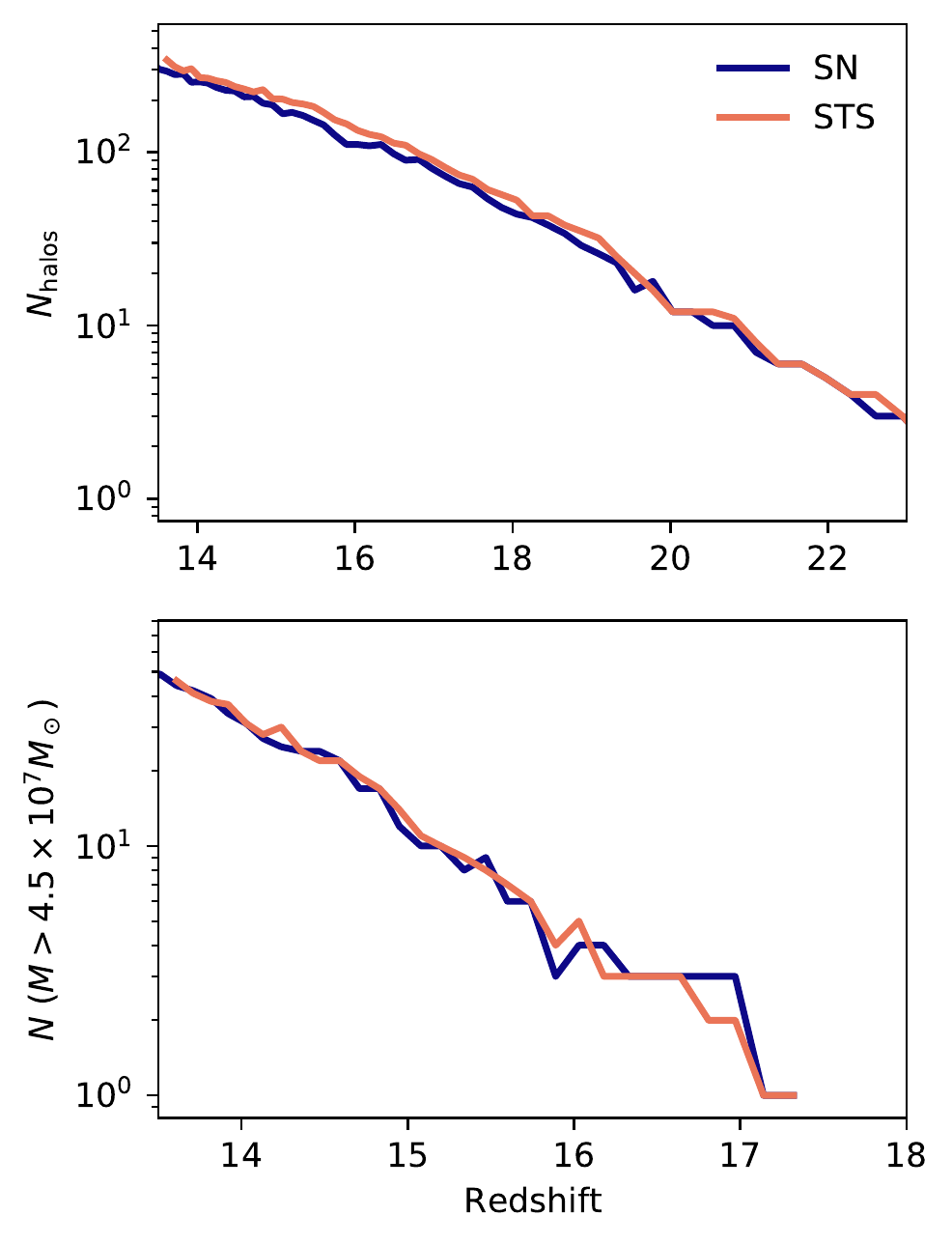}
    \caption{Counts of star-forming halos comparing \sn and \sts suites. \sn halo counts are $\lesssim0.1$ dex below \sts for many redshifts, indicating a systemic but minor reduction in star-forming halo counts due to using \starnet. The lower panel only counts more massive protogalaxies with $M_h > 4.5\times10^7$ \msun.  The difference between \sn and \sts is absent.}
    \label{fig:sn_halocounts}
\end{figure}
    To further describe the volume SFRD, we present the counts of star-forming halos in \sn compared to \sts in Figure \ref{fig:sn_halocounts}.  There are slightly fewer star forming halos in \sn suite at any given redshift.  This, combined with the SFRD that matches \sts, suggests that the individual halos in \sn undergo similar star formation at similar times, on average. However, since there are fewer halos in \sn, there exist some pristine halos in \sn that are enriched star-forming protogalaxies in \sts. In total, \sts contains 8.6\% more star forming halos than \sn.  However, we can see by the lower panel that any difference between \sn and \sts is absent when only considering more massive protogalaxies with $M_h > 4.5\times10^7$ \msun.
    
\begin{figure}
    \centering
    \includegraphics[width=0.48\textwidth]{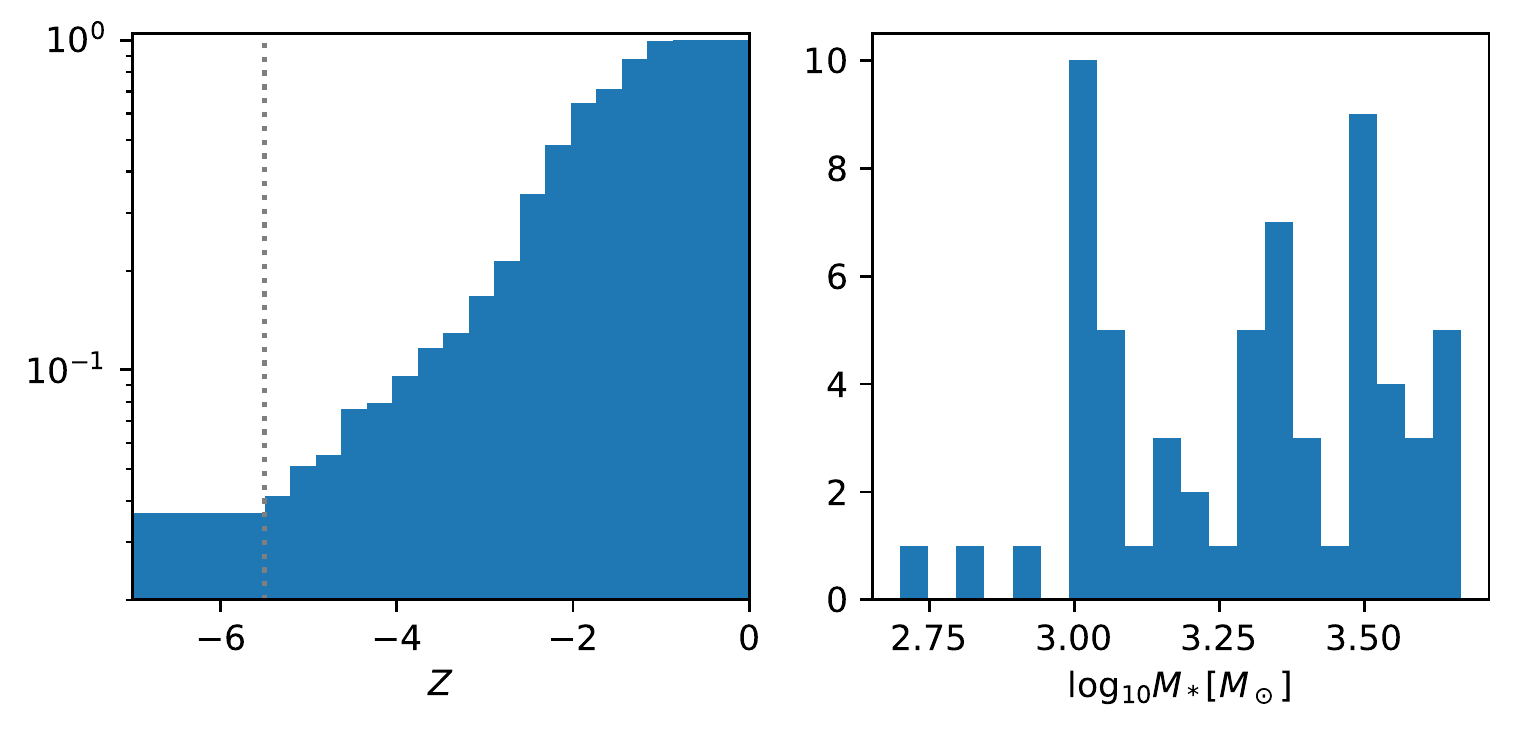}
    \caption{Cumulative MDF of \starss particles formed in sn5l-noZ.  A long tail of low-metallicity clusters have formed with $Z \leq -5.5$, accounting for $\sim4\%$ of stars within the simulation.  We also show the mass of these low-metallicity clusters: they are not low-mass errors, but full $\sim10^3$ \msun clusters that have formed in primordial gas.}
    \label{fig:sn_mdfs}
\end{figure}
    
    In Figure \ref{fig:sn_mdfs}, we show the cumulative MDF of the sn5l-noz simulation.  Of note is the fact that $\sim4$\% of stars in sn5L-noZ formed in gas below $Z_c = -5.5$.  This bulk of stars represents the error that would result from ignoring the requirement of metallicity prior to enriched star formation.  However, there is nothing stopping \starnet from enriching gas well above $Z_c = -5.5$.  We can therefore ask what the effect is when we vary $Z_c$ and ask which stars have formed in error as a function of changing $Z_c$. 
    
\begin{figure}
    \centering
    \includegraphics[width=0.48\textwidth]{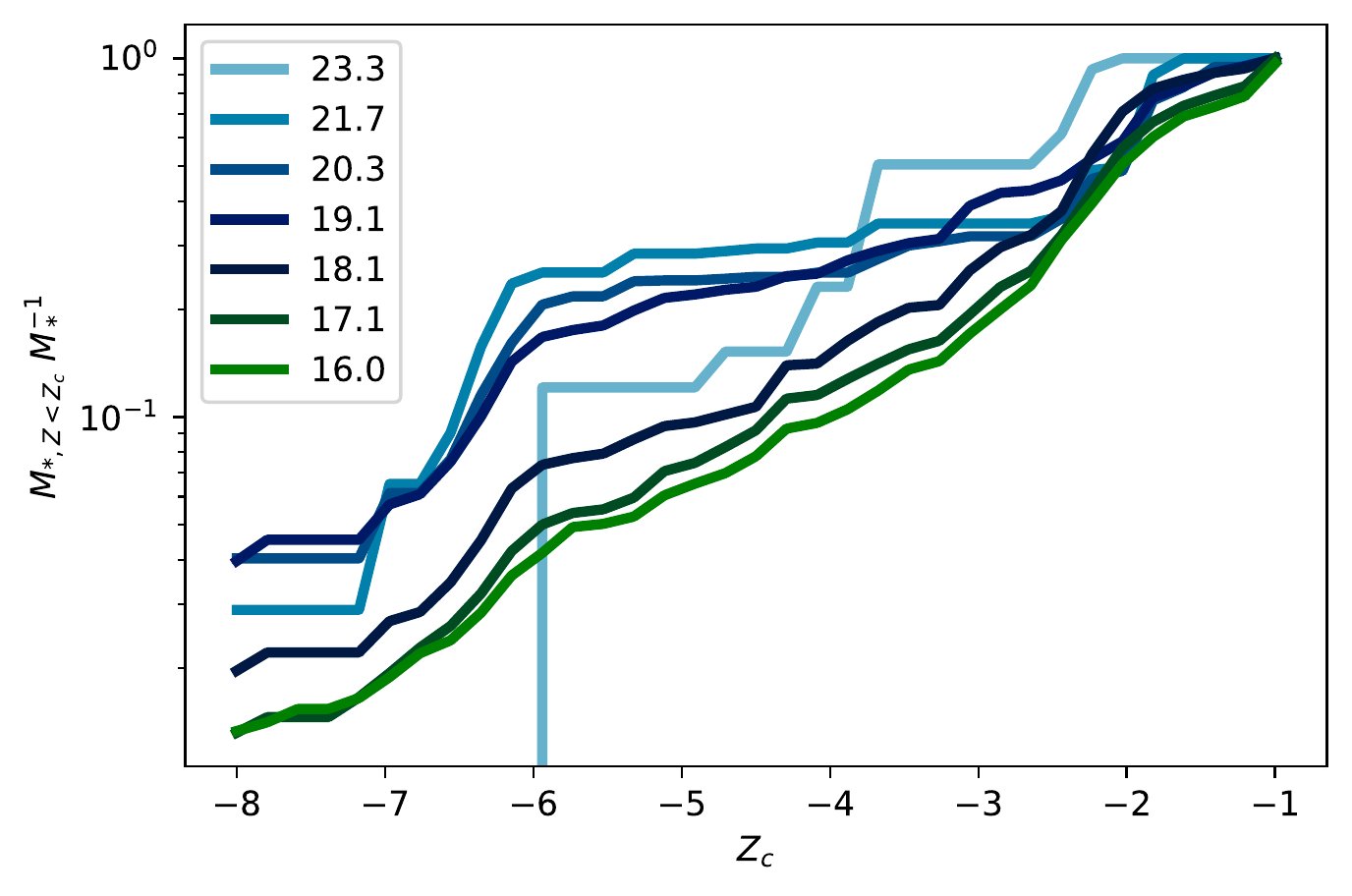}
    \caption{Effect of metallicity floor on stellar mass errors.  The fraction of stars forming with $Z<Z_c$ for varying $Z_c$ are shown at various redshifts. }
    \label{fig:fz_zc}
\end{figure}
    Figure \ref{fig:fz_zc} shows the fraction of stars that formed with $Z<Z_c$ while varying $Z_c$, at various redshifts.  As redshift increases, the central values of $Z_c$ have increasing error, i.e., stars forming with $Z<Z_c$.  In particular, at $z>18.1$, more than 15\% of stars were formed with $Z<-6$, well below the threshold of $Z=-5.5$ used in the simulation.  This is a compounding error upon further reflection:  since some fraction of stars form with $Z<Z_c$, they also contribute to the enrichment of the surrounding gas, leading to further star formation within the region. However, we are not able to differentiate these second-generation errors in this analysis because the star particles do not retain the source of their metallicity (whether from \pii or \piii sources).   The logical conclusion is that some fraction of stars with $Z>Z_c$ are second generation errors that also should not have formed.  
\subsection{Halo and Galaxy Statistics}
\begin{figure*}
    \centering
    
    \includegraphics[width=0.85\textwidth]{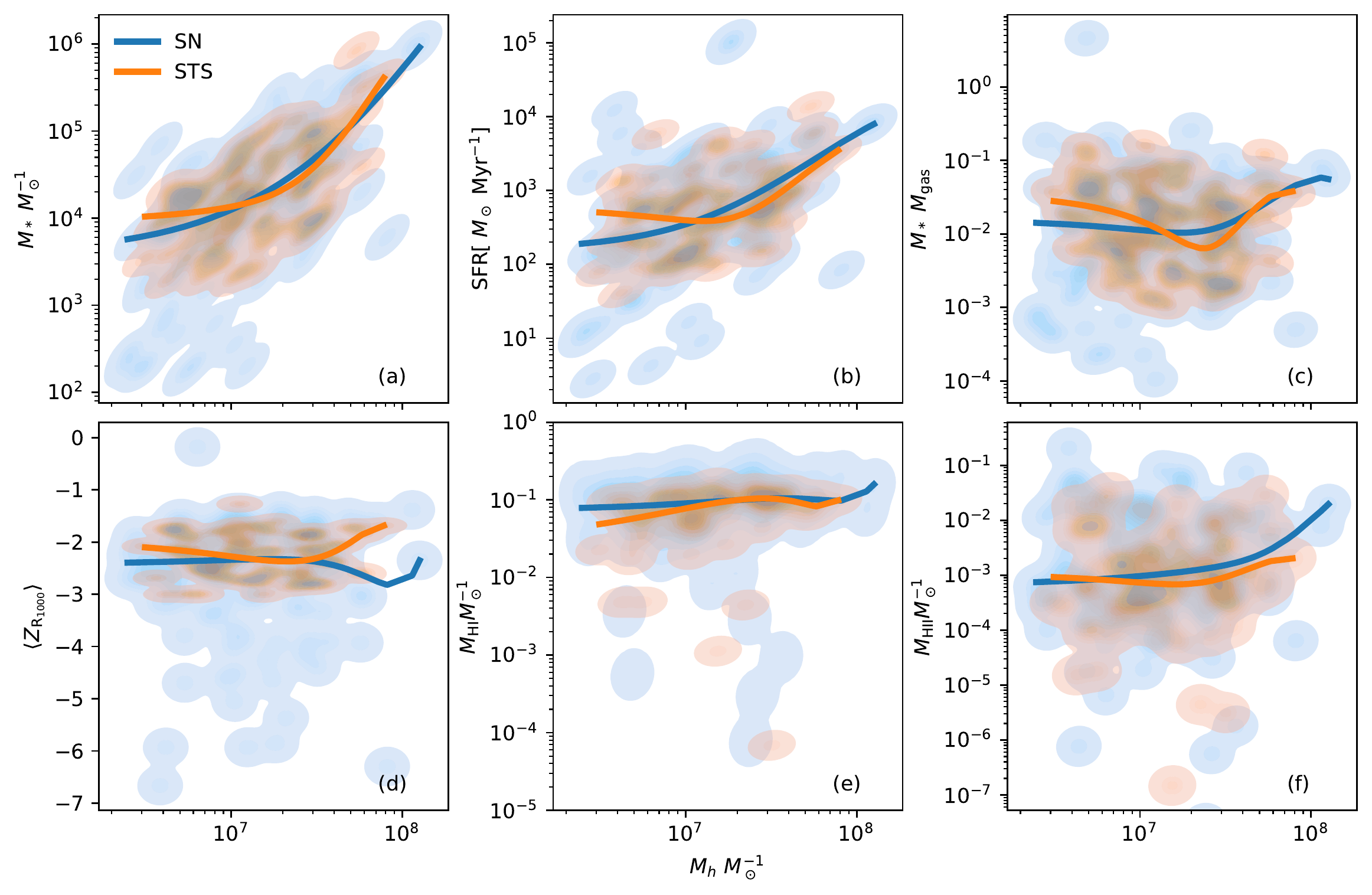}
    
    \caption{Summary statistics for single halos comparing \sts (orange) and \sn (blue) as functions of halo mass.  The full catalog of halos is represented by gaussian kernel density estimation, shown on the background of each plot.  Lines represent a bicubic fit to the 50$^{\rm th}$ quantile of halos, using the degree of fit that minimizes R$_2$ score. }
    \label{fig:single_halos}

    \end{figure*}

    Thus far, the presented statistics focus on a global perspective, disregarding individual halo or star forming regions statistics.  In this final section, we will zoom in and discuss single halos and their proto-galaxies.  To generate our galaxy catalog, we use the {\tt Rockstar} \citep{behroozi2013} halo finder, requiring at least 100 DM particles within the halo.  Each halo is then iterated, and we log various quantities beyond virial radius and mass: stellar mass, averaged historical star formation rate, bolometric luminosity, ionized and neutral hydrogen gas masses, and the mean metallicity of the gas within $R_{1000}$\footnote{Defined as the radius, $R$, at which the mean density of the within $R$, $\bar\rho_h(R)$ satisfies $\bar\rho_h(R) \leq 1000 \bar\rho_b(z)$, for the mean baryon density ($\rho_b$) in the universe at redshift $z$.}.  The bolometric luminosity and optical luminosity is as inherited to the \starss algorithm from FIRE-2.  Figure \ref{fig:single_halos} shows plots of all quantities discussed above.  
At the final redshift, \sn and \sts contain $>10^3$ halos each.  So many individual points would not be legible on any plot, so we represent the full distribution of halos by gaussian kernel density estimation.  To aid in identifying trends in the plotted halos, we also include a bicubic spline fit to the $50^{\rm th}$ quantile of halos.  We can measure the error in the spline fit by comparing the spline prediction $\hat{\mathbf{Y}}$ to the known value ${\mathbf {Y}}$ by $R_2$ score given by:
    \begin{equation}
        R_2({\boldsymbol Y}, \hat {\boldsymbol Y}) = 1 - \frac{1}{N} \frac{\sum_{i=1}^N y_i - \hat y_i}{\sum_{i=1}^N y_i - \bar y}.
    \end{equation}
We then choose the degree of spline fit that maximizes the $R_2$ score;  this prevents overfitting (by simply using high-degree fits) while preserving non-linear behavior that can be achieved with higher-degree fitting.

In general Figure \ref{fig:single_halos} shows that \sn and \sts converge to similar behavior for high-mass halos with $M_h\gtrsim5\times10^7$ \msun.  \sn displays different behavior at lower mass halos, however, showing more low-mass halos with, e.g., low SFR, $M_*/M_{\rm gas}$, and optical luminosity ($L_v$).  Also notable is the extremely lower metallicity gas, $Z_{\rm R_{1000}}$, found in the \sn suite.  The ability of \sn to model low-mass halos is particularly apparent in $M_*$ (Subplot a), where there are several halos with $M_* < 10^3$ M$_\odot$, below the minimum stellar mass observed in \sts.  This suggests that a future application for \sn may be to assist in filling in the faint-end of mass-luminosity relationships in simulations.  Similar behavior is seen in $f_*$ (Subplot c); the low-mass halos with $M_h < 6.8\times10^6$ \msun have significantly reduced star formation efficiency (given as $M_*/M_g$).  \sn also increases the diversity in metallicity of dense regions in star forming halos as shown in Subplot d: while most halos in \sn have similar $\langle Z_{R_{1000}}\rangle$ as \sts, \sn allows for low-metallicity cores to exist, providing an avenue for metal-poor star formation even in previously enriched star forming halos.

One motivating factor to develop \sn was to model pristine halos that could collapse to supermassive black holes, as seen in \cite{regan2017}, without the high resolution requirements of the \textit{Renaissance Simulations} \citep{xu2013} where such halos were found.  To determine if \sn can in fact model these regions, we search \sts for star forming halos and determine whether the corresponding halo in \sn is forming stars.  These ``error halos'' in \sts represent sites where the pristine gas could have ideally formed supermassive black holes, but also could identify large reserves of pristine gas that can fuel star formation in neighboring protogalaxies at later times.

\begin{figure}
    \centering
    \includegraphics[width=0.48\textwidth]{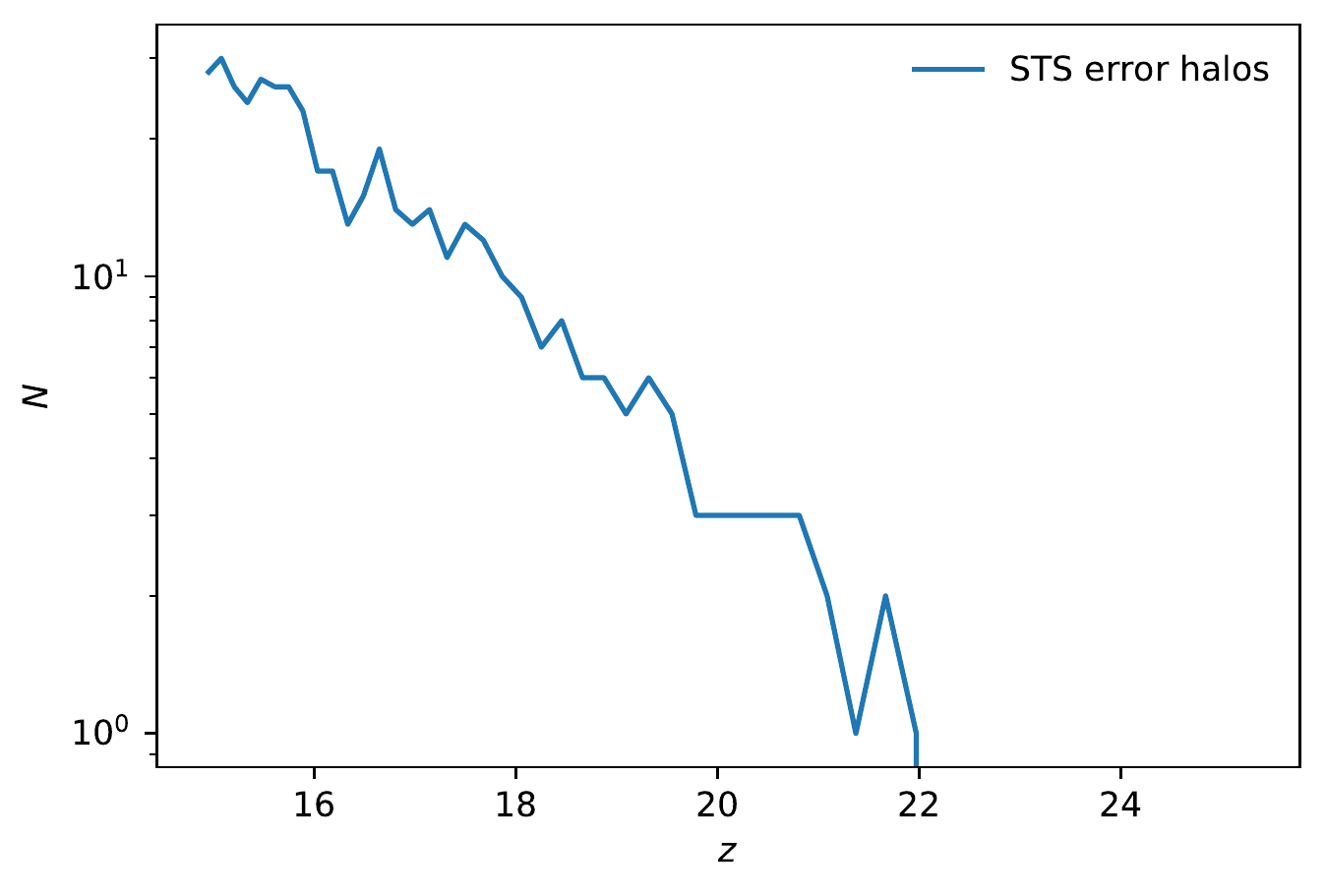}
    \caption{Counts of halos forming stars in \sts where the corresponding halo in \sn has no stellar mass as a function of redshift.}
    \label{fig:error-halos-cnt}
\end{figure}
\begin{figure*}
    \centering
    \includegraphics[width=0.95\textwidth]{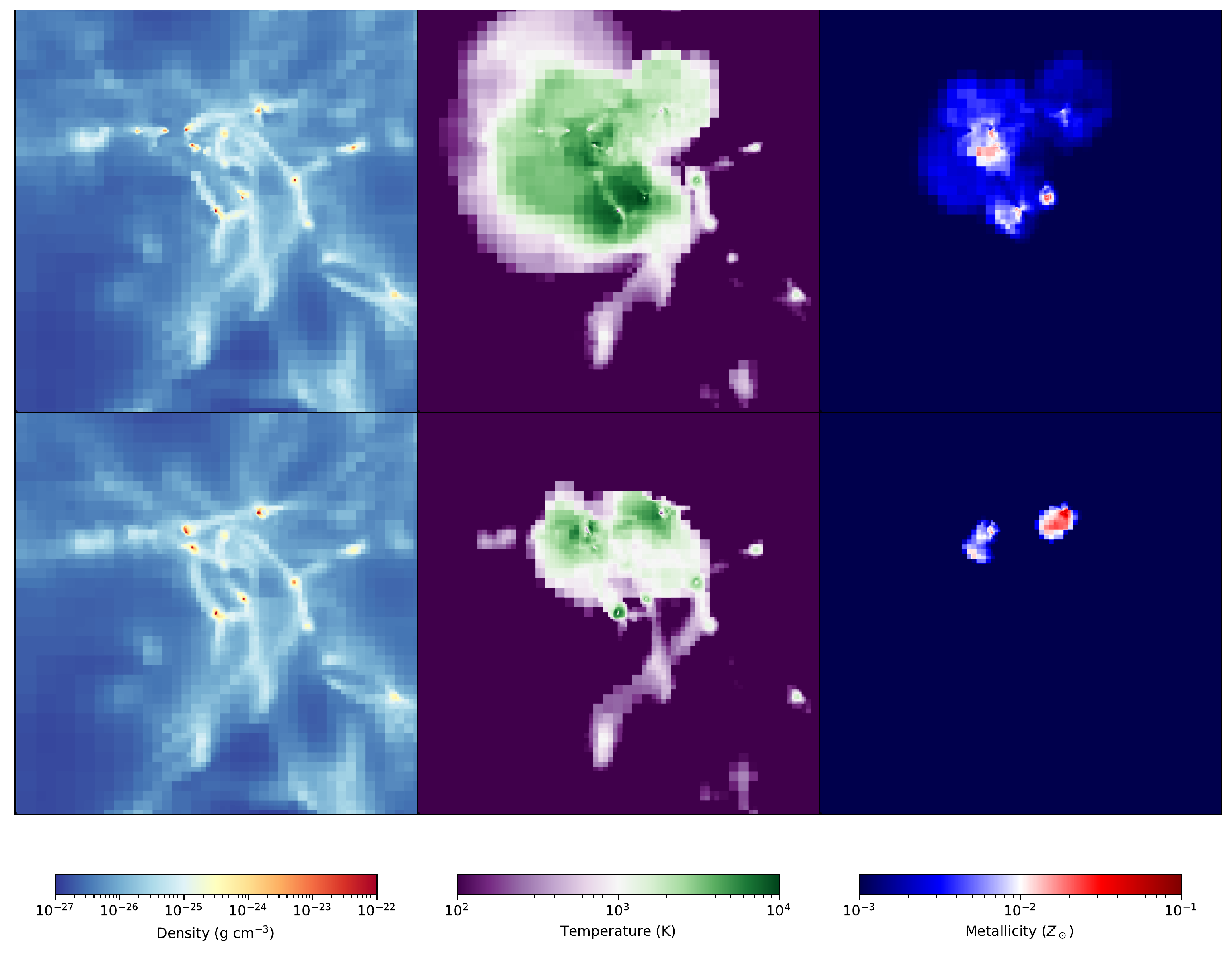}
    \caption{An exemplary halo that is forming stars in \sts (top), but has no stellar mass in \sn (bottom).  The halo of interest is centered in all plots and has $M_v = 8.96\times10^7$ \msun. Metallicity includes metals from all \starss particles and StarNetRuntime depositions.}
    \label{fig:exemp-halo}
\end{figure*}

As shown in Figure \ref{fig:error-halos-cnt}, the number of error halos, i.e., those that are forming stars in \sts but not in \sn, is increasing with decreasing redshift.  While interesting, the {\tt StarFind} component of \starnet is known to have a ``phase'' error, predicting some star forming regions $\lesssim 35$ Myr early, or others $\lesssim 10$ Myr late (W21).  Identifying halos that are truly erroneous would require identifying those halos that are non-star forming in \sn over periods of time $> 30$ Myr, in order to reduce the possibility that the error halo is simply due to phase error. To that end, we analyzed which error halos have existed for the longest times in SN; there are three halos at the final data output, $z=14.95$, that have been forming stars in STS for $\gtrsim 30$ Myr and are not enriched by metals from either \piii or \pii sources.  A prototypical example is presented in Figure \ref{fig:exemp-halo}. In this example, the halo of focus is centered in the frame with $M_h = 8.96\times 10^6$ \msun.  Notably, this halo is in a busy region with many star forming halos nearby (as indicated by the metallicity field sourced from \pii stellar feedback). If a halo such as this one continues without forming stars, it could be a candidate for supermassive black hole collapse or a substantial reservoir of pristine gas for star formation at at later times. Disregarding the phase error, there are 19 pristine halos in \sn corresponding to error halos at the final output, or 26 halos if we include those that have been enriched by \starnet but have not begun star formation. Only one has $M_h > 10^7$ \msun, suggesting that the error halos are less common at higher mass.

\subsection{Observational quantities within \starnet}

The following discusses the effect of including \piii via \starnet from the perspective of observational quantities.  Although the protogalaxies in the \sn we present are far below the observable luminosity of even the James-Webb Space Telescope, these samples may be relatable to ancient protogalactic remnants known as ultra-faint dwarf (UFD) galaxies \citep{simon2019}.  In particular, we will examine the velocity dispersion ($\sigma(\nu))$, stellar metallicity $(Z_*)$, absolute magnitude (M$_{\rm v}$) and neutral and ionized hydrogen masses ($M_{\rm HI}$ and $M_{\rm HII}$).  Each quantity is shown as a function of stellar mass, another easily studied quantity from an observational perspective.  We study these quantities to determine whether the inclusion of \starnet can generate protogalaxies that may evolve into UFDs. 

\begin{figure}
    \centering
    \includegraphics[width=0.48\textwidth]{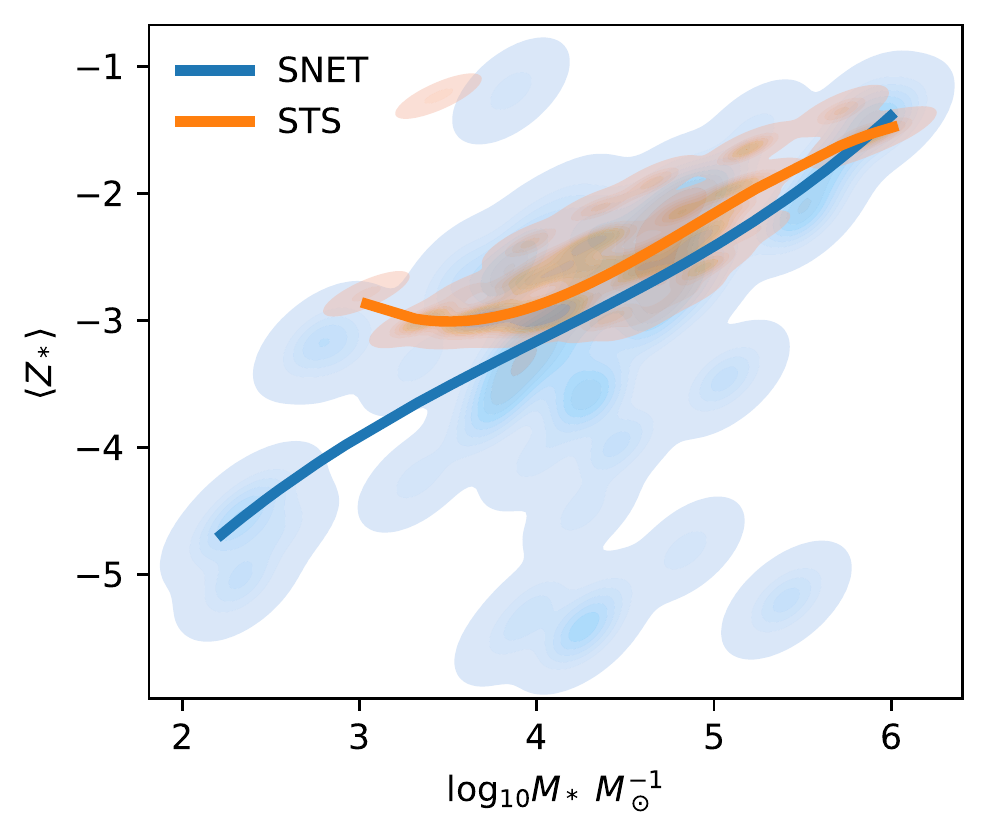}
    \caption{Metallicity of stars as function of halo stellar mass.  The top histogram shows the distribution of stellar masses, while the left side shows the distribution of average stellar metallicity per halo.}
    \label{fig:mstar-zstar}
\end{figure}
Figure \ref{fig:mstar-zstar} shows average stellar metallicity as a function of halo stellar mass.  This dataset continues the trend that \sn generates more diverse behavior than \sts, allowing for exceptionally low-metallicity systems to form at lower stellar mass than is observed in \sts.  The 50$^{\rm th}$ quantile fit shows an increasing trend correlated to stellar mass, however, there is significant scatter allowing for both high and low metallicity examples.  Particularly in \sn, there are middling mass protogalaxies with exceptionally low metallicity.  There is also a much more exaggerated low-metallicity, low-stellar-mass tail present in the \sn suite that is not present in \sts.  

Early results from James Webb Space Telescope (JWST) have found several galaxies at $z\sim 8$ with low metallicity, albeit at higher stellar mass than the protogalaxies observed in this work. \citep{curti2022}.  
\begin{figure}
    \centering
    \includegraphics[width=0.48\textwidth]{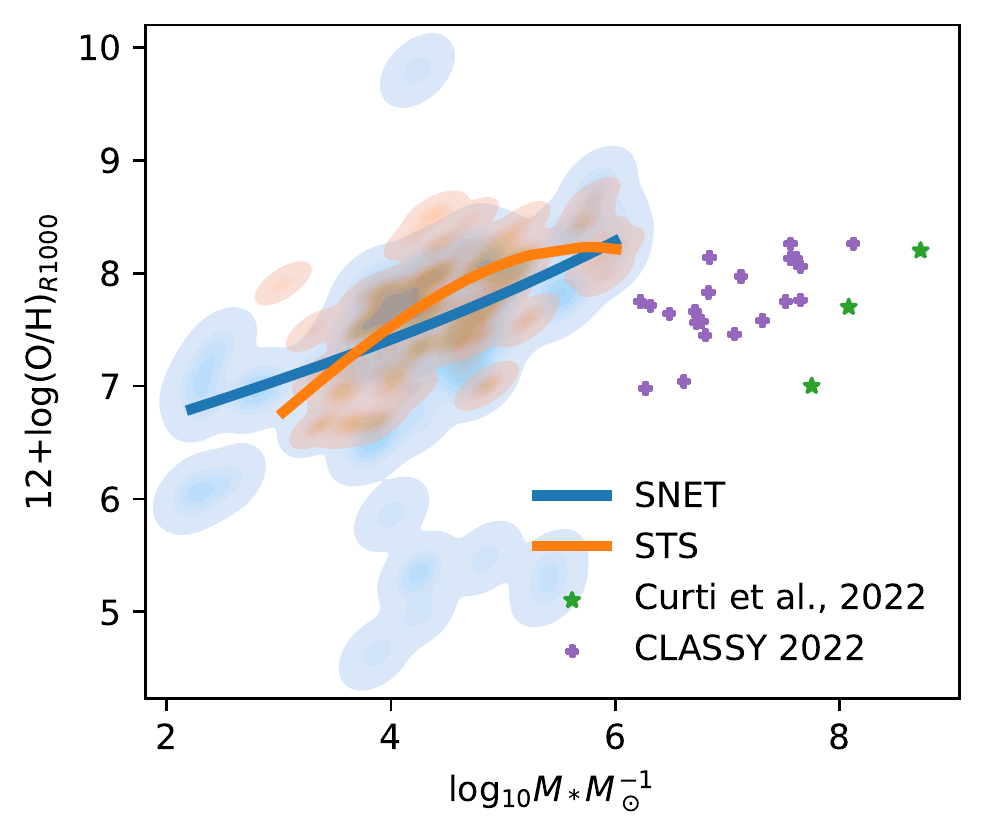}
    \caption{O/H as function of stellar mass.  Green stars represent the level of metallicity observed in \citet{curti2022}, although their observed galaxies had much higher stellar mass ($10^{7.5}< M_*/M_\odot < 10^8$).  We also annotate a low-stellar mass subset of samples obtained from the CLASSY database in purple \citep{berg2022}}
    \label{fig:mstar-zgas}
\end{figure}
Figure \ref{fig:mstar-zgas} shows the gas-phase metallicity ([O/H]), as in \citet{curti2022}, using the approximation that oxygen comprises 35\% of the mass-weighted metallicity within $R_{500}$ of the halo \citep{torrey2019}.  Since our protogalaxies are of much lower stellar mass, we show the level of the observed metallicity by green stars, however, the stellar mass of their observations are $> 10^{7.5}$ \msun.  Nonetheless, the inclusion of \sn again allows more diverse behaviour in our simulated protogalaxies, so that their low metallicity regime overlaps with that of the observations from JWST. The inclusion of \starnet also enables the range of metallicity seen in the CLASSY database \citep{berg2022}, however, most samples there are also much higher in stellar mass than the protogalaxies observed in \starnet's small volumes.  In order to expand on this relationship, \starnet will need a more efficient implementation that can explore larger volumes and higher-mass protogalaxies.

\begin{figure}
    \centering
    \includegraphics[width=0.48\textwidth]{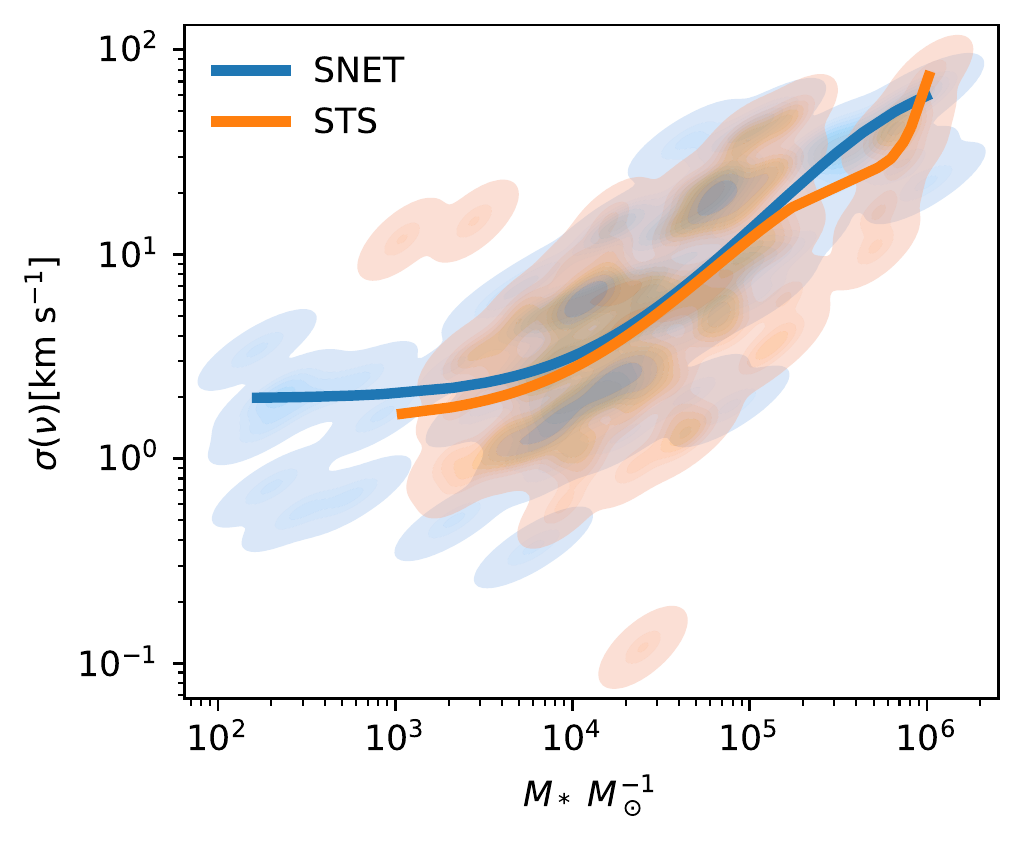}
    \caption{Velocity dispersion as function of stellar mass.  }
    \label{fig:mstar-sigmav}
\end{figure}
\begin{figure}
    \centering
    \includegraphics[width=0.48\textwidth]{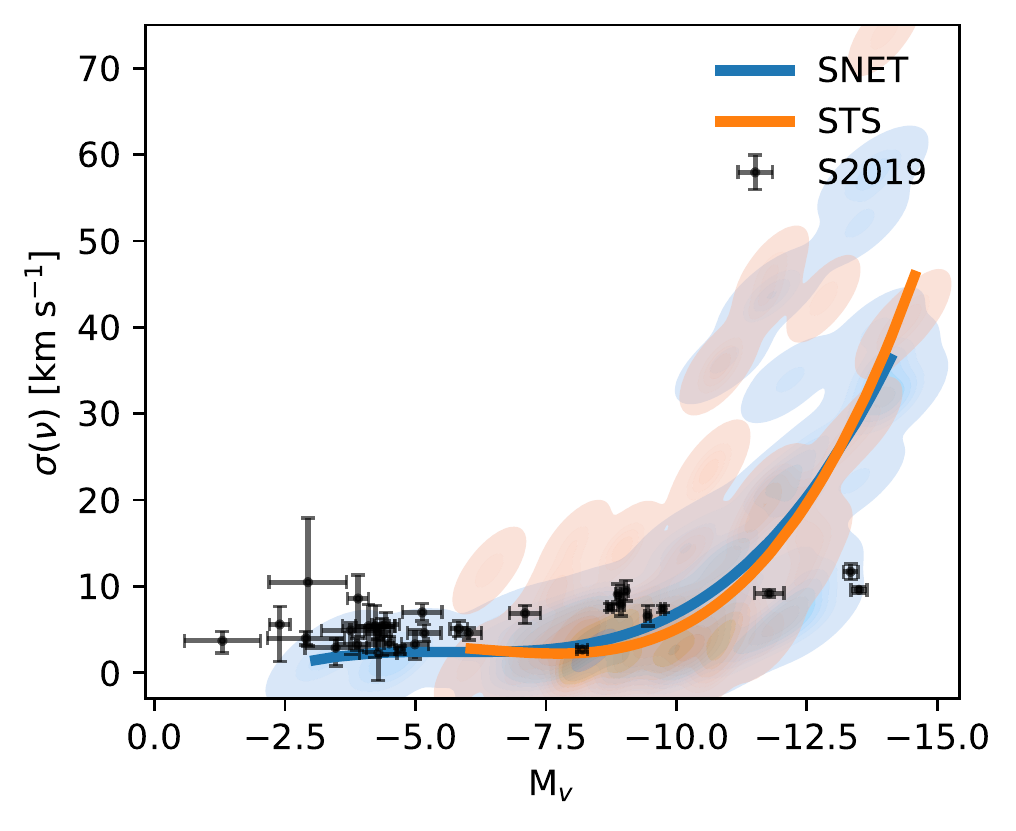}
    \caption{Velocity dispersion as function of absolute magnitude. Also shown are observational data of ultra-faint dwarf galaxies (see text for references).}
    \label{fig:mv-sigmav}
\end{figure}
Figure \ref{fig:mstar-sigmav} shows the radial velocity dispersion as a function of stellar mass.  All samples above the 50$^{\rm th}$ quantile have $\sigma(\nu) > 2$, where most UFDs are found \citep{simon2019}.  This sample does have a significant population with lower dispersion, typically at low stellar mass ($M_* < 10^5$ \msun).  
Figure \ref{fig:mv-sigmav} builds on the study of $\sigma(\nu)$ by showing $\sigma(\nu)$ as a function of absolute magnitude (M$_{\rm v}$). UFDs typically occupy the region with M$_{\rm v} > -8$: this regime is probed by both \sn and \sts, however there is a higher magnitude ($-2 < M_{\rm v} < -5$) sample of halos with low velocity dispersion that is only present in \sn.  

It may be possible that the inclusion of \starnet allows the simulation to capture the observed behavior of low-mass dwarf galaxies more effectively than simulations employing metallicity floors.  To explore \sn and \sts compared to observations, we include observational data along with our simulated data.  The observations were compiled by \cite{simon2019}, and include the works of many authors \citep{simon2007, walker2009a, walker2009b, bechtol2015, koposov2011, koposov2015, crnojevic2016,simon2015, li2017, li2018,simon2011, torrealba2018, mateo2008, willman2011, spencer2017, collins2017, torrealba2016, caldwell2017, kirby2015, koch2009, majewski2003, bellazzini2008, kimd2015, kimd2015b, torrealba2016b, walker2016, munoz2018}. \sn extends the high-magnitude tail past $M_{\rm v} > -6$, allowing the potential coverage of many observations with $\sigma(v) < 10$ and $M_{\rm v} > -6$, where the bulk of UFDs reside.  Further evolution of the \sn suite will be interesting to see if the high-magnitude, low $\sigma(\nu)$ region becomes more well-represented.
\begin{figure*}
    \centering
    \subfloat[]{
        \includegraphics[width=0.48\textwidth]{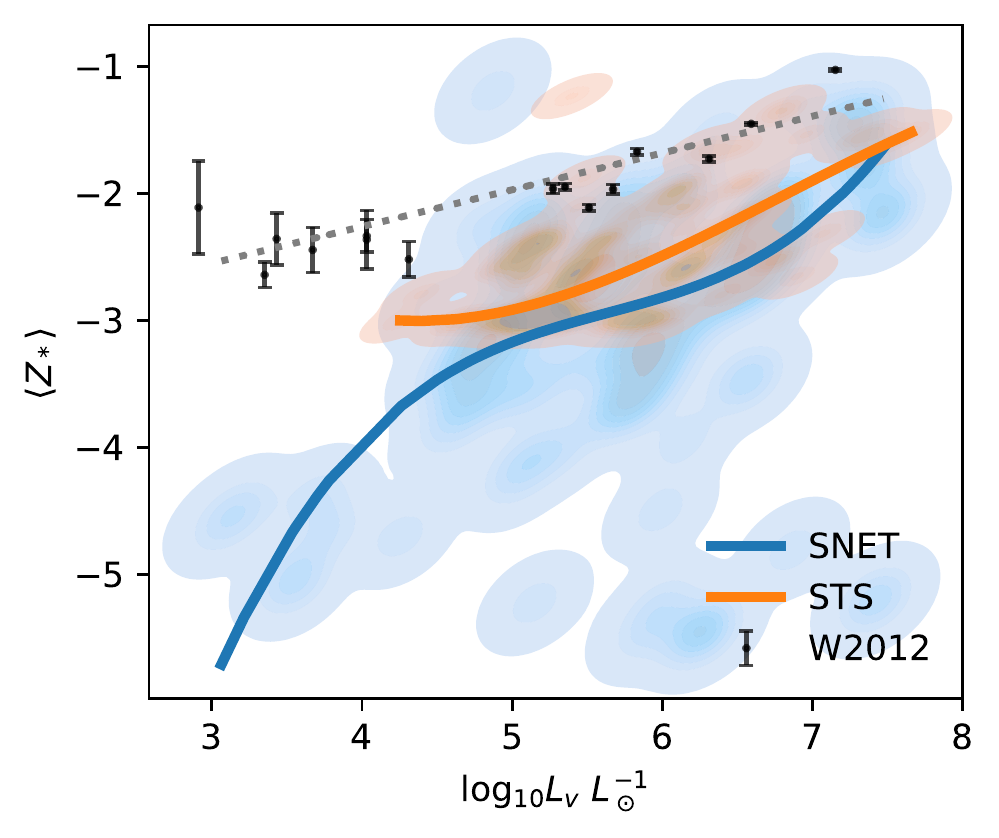}
        \label{fig:lv-zstar}}
        \subfloat[]{
        \includegraphics[width=0.48\textwidth]{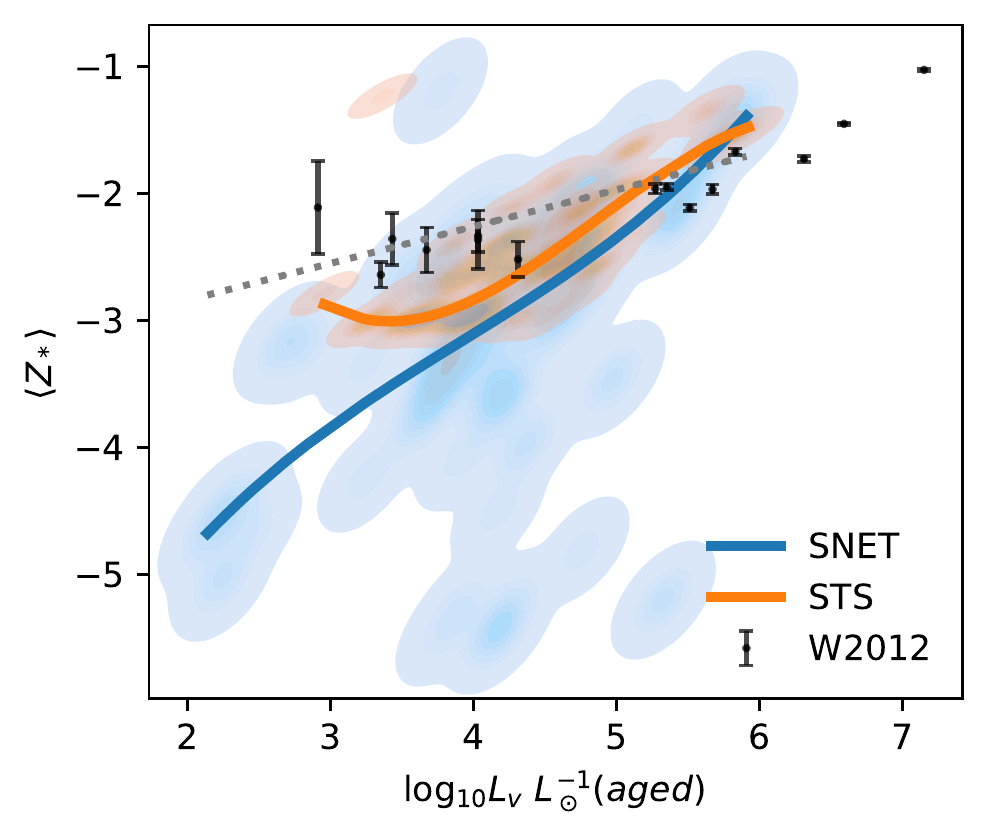}
        \label{fig:lv-zstar-old}}
    \caption{Mean stellar metallicity as function of luminosity.  Panel \ref{fig:lv-zstar} shows the total halo luminosity for the stellar ages and masses present within the virial radius.  Panel \ref{fig:lv-zstar-old} shows the same quantity, if we suppose the stellar population has aged by 4 Gyr.  The expected luminosity-metallicity relationship (Equation \ref{eqn:metallicity}) is also shown (gray hashed line).  Local group dwarfs (black points) are also shown \citep{willman2012} with their corresponding error estimates.}
    \label{fig:lv-zstar-tot}
\end{figure*}
We finish our study of observational quantities with Figure \ref{fig:lv-zstar}, showing stellar metallicity as a function of luminosity. Also shown is the metallicity-luminosity relation given by 
\begin{equation}
    [{\rm Fe/H}] = -1.68 + 0.29 \log\bigg(\frac{L_{\rm v}}{10^6L_\odot}\bigg),
    \label{eqn:metallicity}
\end{equation}
although we use total mass-weighted metallicity $Z$ as a proxy for [Fe/H]\footnote{This assumption holds true for solar metallicity stars, but will break down for, e.g., carbon-enhanced metal poor stars whose abundances do not resemble solar.} \citep{simon2019}.  Both \sn and \sts have a broad, 3-5 dex spread of metallicity that is not represented by Equation \ref{eqn:metallicity}, however, both suites have samples that overlap the analytic model.  As has been the trend, the inclusion of \sn permits a much broader range of halo observables, increasing the metallicity range from $\sim3$ dex to $\sim5.5$ dex.  

To compare \sn to observations, we include dwarf and ultra-faint dwarf (UFD) observations of the Local Group \citep{willman2012} (W2012).  The dwarf galaxies are well represented by both \sn and \sts, however, both simulation suites tend toward lower metallicity and do not duplicate the low-luminosity end of observational data.  However, \sn does have a low-luminosity tail that extends almost as low as observations, albeit with much lower mean metallicity.  The time evolution of this low-luminosity tail will be of great interest in studies where simulations approach reionization, $z\lesssim 6-7$.

Since our sample of halos is taken from $z=14.95$, one may ask what the relationship would look like for a halo that stopped forming stars at that point and aged to more modern times.  Since UFDs are thought to have ceased star formation very long ago, $z\sim 6-7$, then this may serve as a rough approximation of how some halos could age within the \sts and \sn framework.  In Figure \ref{fig:lv-zstar-old}, we have plotted the luminosity-metallicity relationship, however, the luminosity of the stellar population has been aged by 4 Gyr.  The distribution of luminosity has much more overlap with the UFD observational points, although a lack of metallicity evolution means that neither \sts or \sn match the expected slope of Equation \ref{eqn:metallicity}.

\section{Discussion}

In this work we have developed both a resolution-intelligent, physically-motivated feedback method for stellar sources in \enzo (\starss), and a new method to initialize the metallicity field of astrophysical simulations (\starnet).  \starss has shown success as a resolution-intelligent feedback algorithm, but still struggles with resolution effects regarding star formation.  If cooling dynamics are substituted for force resolution, i.e., using $Z_{\rm f} = -3$ or using \starnet to seed the metallicity field, much of the resolution dependence is resolved.  That said, creating a truely resolution-intelligent star formation algorithm will require a new paradigm of star formation recipes;  the current criteria are \textit{always} resolution sensitive to some degree, because the force resolution of the simulation is intrinsically related to the spatial resolution.  One possibility is to use super-resolution techniques \citep[e.g.,][]{kappeler2016} as a surrogate model of high-resolution gas dynamics in low-resolution regions, but this is left to future work.  With \starss current implementation, lowered resolution acts to delay star formation to later times, but maintains a similar SFR once star formation begins.  This suggests that at lower redshift, the effect may be disregarded, depending on the focus of the simulation. 

\starnet in its current state is a proof-of-concept work.  It uses the inline Python capability of \enzo, which severely limits its scalability.  Despite this restriction, we have simulated a significant portion of the canonical \piii era ($10 < z < 30$) in two 17.56 Mpc$^3$ simulation volumes.  Using \starnet as a surrogate for \piii star formation and feedback enables substantial speedup, even in this unoptimized state.  Using the \textit{Phoenix Simulations} as a reference, PHX256-2 required $\sim 5600$ node-hours on the TACC-Frontera supercomputer using 6 nodes.  There is significant acceleration using \starnet: sn5l-2-v3 required only 1025 node-hours running on 2 nodes, for a total speedup of $5.46\times$.  It is useful to note that this speedup is the minimum to expect from future implementations of \starnet: incorporating \starnet into \enzoe will enable load balancing and will benefit from the framework being translated to C++. 

The inclusion of \starnet significantly modifies simulation behavior, not from a global perspective, but from the perspective of outliers and rare events.  \sn and \sts show similar SFRD within their respective volumes, but the slight differences in SFRD generate large differences in cumulative stellar mass; \sts overproduces stars by $\sim 20\%$ by $z=15$, despite under-producing stars at early times ($z\gtrsim 21$).  The global similarity between \sn and \sts is reinforced by examining star-forming halo counts, where we observe very similar numbers of star-forming halos in both \sn and \sts.  Any difference in regards to halo counts is largely resolved if we only consider higher-mass halos ($M_h \gtrsim 4.5\times 10^7$ \msun).  The difference in halo counts, resulting in those forming stars in \sts but not in \sn, is increasing with decreasing redshift, suggesting that this is an error that would become more significant at lower redshifts.

One simulation in this work was designed solely to identify which stars would form in error if we removed the $Z_c$ criterion for star formation.  We find that this simulation overproduces stars by $\sim4\%$ at $z\sim 16$.  Although the difference is minor, this effect is more pronounced at higher redshift, with errors as high as 30\% at $z\sim23$.  The error also increases if we increase $Z_c$, with as much as 15\% of stars having $Z<-3$ at $z\sim 16$.  The errors we note here are likely compounding: a single cluster formed in error can enrich neighboring gas, leading to more star formation that is also in error.

The net effect of including \starnet to model the initial metallicity field is the increased range of behaviors that can be observed in Figures \ref{fig:single_halos}-\ref{fig:lv-zstar-tot}.  Particularly when we include observational data of dwarf galaxies in Figures \ref{fig:mv-sigmav} and \ref{fig:lv-zstar-tot}, the inclusion of \starnet leads to halos whose characteristics overlap more substantially with observational data.  While the shown observations are near to $z=0$, there is suspicion that the UFDs have been essentially static since high-redshift \citep{simon2019} from a star formation and baryonic gas perspective, suggesting that they may be relics of a high-redshift universe.  If so, then there should be a redshift at which our halo distributions begin to overlap with the observational data.  This may not happen until reionization, $z\sim 7$, since many of these halos do not have enough gas to self-shield from external ionizing radiation.  Such a simulation will be attainable once \starnet has been incorporated into \enzoe.  For now, if we synthetically age the population of dwarf galaxies in \sts, the resulting $L_{\rm v}$-$Z_*$ relationship overlaps much more with observations of UFDs.  This quick and dirty comparison suggests that the age and star formation cutoff will have a significant impact on modeling UFDs, and modeling them with \starnet enables a much lower $L_{\rm v}$ cutoff at $L_{\rm v}\sim 10^2$ L$_\odot$.  Further reinforcing the importance of the dynamic range in metallicity achieved by incorporating \starnet, protogalaxies in \sn show the full range of stellar metallicity as observed in early results from the JWST, where $z\sim8$ galaxies have been identified with $-5<Z<-3.75$ \citep{curti2022}.  Although the observed galaxies have higher stellar mass than our simulated protogalaxies, a larger simulation domain and incorporation of \starnet into \enzoe will enable us to explore galaxies in the stellar mass range of the JWST observations.

\starnet has also shown the ability to model rare events, such as a halo that, while forming stars in \sts, has not been pre-enriched by a \pa and so remains pristine in \sn.  These halos are uncommon, with only nine examples at the final redshift of \sn and only three examples that have existed for $>35 Myr$. Protogalaxies in \sts that have yet to begin star formation in \sn are more common, with 8.6\% (26 examples) of protogalaxies in \sts having no twin in \sn.  However, these samples may be extremely important to later evolution, serving as pristine reservoirs of gas to fuel star formation or as potential sites of supermassive black hole formation.  Optimizing and integrating \starnet in future \sn simulations with larger volume to lower redshift will better elucidate the evolution, lifetime, and impact of these pristine halos.

\bigskip
\noindent
This research was supported by National Science Foundation grant CDS\&E grant AST-2108076 to M.L.N. The simulations
were performed using \enzo on the Frontera supercomputer operated
by the Texas Advanced Computing Center with LRAC allocation AST20007 and the Expanse supercomputer operated by the San Diego Supercomputer Center for XSEDE with XRAC allocation AST200019.  Simulations were performed with \enzo \citep{enzo, brummel-smith2019} and analysis with {\tt YT} \citep{turk2011}, both of which are collaborative open source codes representing efforts from many independent scientists around the world.

\bibliographystyle{aasjournal}
\bibliography{bib}
\end{document}